\documentclass[12pt]{article}
\usepackage{eurosym}
\usepackage{float}
\usepackage{graphicx}
\usepackage[utf8]{inputenc}
\usepackage[T1]{fontenc}
\usepackage{subcaption}
\usepackage{indentfirst}
\usepackage[margin=0.6in,nomarginpar]{geometry}
\usepackage[final]{hyperref}
\usepackage{amsmath}
\usepackage{hyperref}
\usepackage{cite}
\usepackage{subcaption}
\usepackage{caption}
\usepackage{amssymb}
\usepackage{multirow} 
\usepackage[table]{xcolor}
\usepackage{orcidlink}
\usepackage{color}
\usepackage{hyperref}
\usepackage[utf8]{inputenc}

\hypersetup{
colorlinks=true,
linkcolor=blue,
citecolor=blue,
filecolor=magenta,
urlcolor=blue
}
\begin{document}

\title{Thermal aspects and particle dynamics of Euler-Heisenberg AdS black hole in 4D Einstein Gauss-Bonnet gravity}
\author{B. Hamil$^{a}$ \orcidlink{0000-0002-7043-6104} Faisal Javed$^{b,c}$ \orcidlink{0000-0002-7043-6104} \thanks{%
hamilbilel@gmail.com\\faisaljaved.math@gmail.com} \\
$^{a}$Laboratoire de Physique Math\'{e}matique et Physique Subatomique,\\
Facult\'{e} des Sciences Exactes, Universit\'{e} Constantine 1, Constantine,
Algeria.\\
$^{b}$Department of Physics, Zhejiang Normal University, Jinhua 321004,\\ People's Republic of China\\
$^{c}$Research Center of Astrophysics and Cosmology, Khazar University,\\ Baku, AZ1096, 41 Mehseti Street, Azerbaijan}

\date{\today }
\maketitle

\begin{abstract}
We construct charged  AdS BH solutions in four-dimensional Einstein-Gauss–Bonnet gravity coupled to Euler–Heisenberg nonlinear electrodynamics and investigate their physical properties. The modified field equations admit black hole solutions whose horizon structure is significantly affected by higher-curvature and nonlinear electromagnetic corrections, allowing for multiple horizons depending on the model parameters. In the extended phase space, where the cosmological constant is interpreted as thermodynamic pressure, we analyze the thermodynamic behavior and show that both the Gauss–Bonnet coupling and the Euler–Heisenberg parameter induce notable modifications in the equation of state, critical behavior, and thermal stability. Interpreting the black hole mass as enthalpy, we study the Joule-Thomson expansion and determine the inversion temperature and pressure, demonstrating that higher-curvature and nonlinear electrodynamic effects substantially influence the cooling and heating regions. Finally, we examine timelike geodesics and show that Gauss–Bonnet corrections significantly modify the effective potential, orbital stability, and particle motion in the strong-field regime.
\end{abstract}

\section{Introduction}
\label{sec.1}
Nonlinear electromagnetic extensions of the Reissner-Nordstr\"{o}m solution in Einstein–Maxwell theory have been extensively studied in recent years. A particularly important example is provided by the
gravitating Born-Infeld (BI) theory \cite{Born}. Early studies of
charged black holes (BHs) coupled to nonlinear electrodynamics (NLED) date back to
the 1930s \cite{Hoffmann,Infeld}. Subsequent developments in string theory
and D-brane physics revealed that Abelian and non-Abelian BI-type
Lagrangians naturally emerge in the low-energy limit \cite{Fradkin}, thereby
renewing interest in nonlinear electromagnetic actions. Within this
framework, BH
solutions of Einstein-Born-Infeld gravity were obtained in \cite%
{Garcia,Demianski}.

Nonlinear extensions of Reissner–Nordström solutions in Einstein-Maxwell theory has drawn considerable attention while discussing the shortcomings of linear electrodynamics at the strong-field regime. Among these extensions, the gravitating Born–Infeld (BI) theory \cite{Born} is particularly interesting as BI has a strong physical motivation as he was the first to propose its application to regularizing the divergent self-energy of point charges. The study of static electrically charged BHs in the context of NLED goes back to the pioneering work of the 1930's \cite{Hoffmann,Infeld} in the absence of a BH where a horizon was understood. It was noted that the nonlinear correction may trigger a strong effect on the structure of the solutions' near horizon. With the progress of string theory and D-brane theory, it was possible to extend those theories and provide motivation for Abelian and non Abelian BI type Lagrangians \cite{Fradkin} which substantially attracted attention on the  nonlinear electromagnetic theory in gravity  context. In the context of the aforementioned references, asymptotically flat static and spherically symmetric BH solutions of Einstein- Born-Infeld gravity \cite{Garcia,Demianski  } were systematically studied and found to possess deviations from the Reissner Nordstrom geometry to non-Neilsian electrodynamics.

A variety of NLED models admitting static and
spherically symmetric solutions have also been investigated over the past
decades. These include models with Lagrangians given by general functions of
the electromagnetic invariants $F_{\mu \nu }F^{\mu \nu }$ \cite%
{Alonso,Rubiera}, logarithmic modifications of the Maxwell invariant \cite%
{Soleng}, and generalized nonlinear Lagrangians \cite{Oliveira} that
reproduce both the BI theory and the weak-field limit of the
Euler-Heisenberg (EH) effective Lagrangian \cite{Heisenberg}. BH
solutions arising from the coupling of gravity to NLED
in the weak-field EH limit, interpreted as a low-energy
approximation of the BI theory, were analyzed in \cite{Yajima}. Considerable
effort has also been devoted to the construction of regular BHs in gravitating NLED \cite%
{Beato,Ayon,Burinskii}, with their distinctive properties discussed in \cite%
{Sulaman,Bilel,Eslam,Phukon,Davood,Dmitri,Hassaine,Hideki,Angel,Bokulic,Dehghani,Halilsoy,Dayyani,Habib,Gabriel,Hamil}.

In the framework of Dirac's electron–positron theory, nonlinear electromagnetic fields were first modeled by Heisenberg and Euler \cite{Heisenberg}. Later, these non-perturbative one-loop effective Lagrangians were reformulated by Schwinger within the context of QED \cite{Schwinger}. The effective theory obtained describes vacuum polarization, whose imaginary part corresponds to the vacuum decaying via electron–positron pair production. If the electric field value is above the critical value, which is given by $E_{c}=\frac{m^{3}c^{3}}{e\hbar }$ the vacuum can be polarized by the spontaneous pair creation \cite{Heisenberg,Schwinger}. For many years, a lot of theories and experiments have been oriented toward pair creation from QED vacuum and vacuum polarization due to external electromagnetic fields \cite{Bilel1,Bilel2,Bilel3,Gitman1,Gitman2,Gitman3,Gitman4}. As a
fundamental and experimentally well-tested theory, QED provides an accurate
description of electromagnetic interactions, making the investigation of QED
effects in BH physics particularly important. Consequently, the
EH effective Lagrangian plays a central role in the
construction and analysis of generalized BH solutions discussed
above.

Lovelock theories \cite{Lovelock1,Lovelock2}, which incorporate higher-order curvature terms, provide a natural generalization of Einstein's general relativity to higher-dimensional spacetimes, while preserving the second-order nature of the field equations (FEs) for the metric. So Lovelock gravity bypasses some challenges that are typical in other higher-derivative theories of gravity, where the equations of motion are of the fourth order or higher, and where linear perturbation analyses result in the emergence of ghost instabilities, among other things. More specifically, Einstein-Gauss-Bonnet (EGB) gravity, which is a second-order Lovelock theory, consists of a Lagrangian containing the Einstein-Hilbert term with a cosmological constant, and the quadratic Gauss-Bonnet (GB) term. EGB gravity has been the subject of a lot of research, in part because it is the low-energy effective theory of strings propagating in curved spacetime \cite{Wiltshire1,Wiltshire2}. In this context, BH solutions that are static and spherically symmetric were found by Boulware and Deser \cite{Boulware}. In 4-dimensions, the GB contribution is topological and does not modify the gravitational equations of motion in the absence of matter couplings. \cite{Odintsov,Oikonomou}.

More recently, a four-dimensional version of EGB gravity was introduced by
Glavan and Lin \cite{Glavan} via a regularization scheme, in which the GB
coupling $\alpha $ is rescaled as $\frac{\alpha }{D-4}$ in $D$ dimensions
and the limit $D\rightarrow 4$ is taken at the level of the NLED.
It is worth noting that this regularization scheme has been the subject of
ongoing debate, and several shortcomings have been pointed out in the
literature \cite{Shu,Arrechea,Mitchell}. Various proposals have been put
forward to address these issues \cite{Lu}, yielding the same BH
metric as in the original formulation \cite{Glavan} and suggesting that it
remains a valid solution. Consequently, the original regularization scheme
may still be employed to investigate EGB gravity coupled to additional
fields and to construct new BH solutions.
Research on BHs in 4-dimensional EGB gravity has attracted considerable attention, and various aspects such as thermodynamics, quasinormal modes, BH shadows, and stellar structure have been extensively studied within this framework \cite{RKumar,SGangopadhyay,GGhosh,Islam,ABiswas,HShah,Heydari,Bouali,Ahmedov,YSekhmani,Birkandan,DMondal,BHamil,GPedro,UPapnoi,Konoplya,PKYerra,WenWei,KJafarzade}.

Four-dimensional EGB gravity provides an effective higher-curvature extension of general relativity that admits well-defined BH solutions while retaining second-order NLED. EH NLED, derived as the one-loop effective action of QED in the weak-field limit, introduces nonlinear electromagnetic corrections associated with vacuum polarization effects. Investigating BHs within this combined framework enables the assessment of the interplay between higher-curvature gravity and nonlinear electromagnetic fields, clarifying their impact on the structure and physical properties of BH spacetimes. This model allows for a detailed analysis of thermodynamic behavior and phase structure, including the Joule-Thomson (JT) expansion and the associated cooling and heating processes. Since both GB and EH corrections are expected to influence the equation of state and thermodynamic response, a joint analysis is required to understand their combined effects on cooling and heating processes.In addition, the examination of test particle motion offers complementary information on orbital stability and strong-field effects induced by these corrections. Motivated by these considerations, we analyze charged AdS BHs in D4 EGB gravity coupled to EH NLED, focusing on their thermodynamics, JT expansion, and geodesic structure.

This paper is structured as follows. Section \ref{sec:2} discusses the action of EGB gravity coupled to EH NLED. We derive the relevant NLED and BH solutions. In Sect. \ref{sec:3}, we explore BH thermodynamics in the extended phase space: equations of state, critical phenomena, and thermal stability. Sect. \ref{sec:4} concerns the JT expansion, where we obtain inversion temperature and pressure and describe the relevant cooling and heating regions. In Sect. \ref{sec:5}, we discuss the motion of test particles along geodesics, with an emphasis on the effective potential, circular orbits, epicyclic frequencies, and the innermost stable circular orbit (ISCO). Our conclusions are stated in Sect. \ref{sec:6}.

\section{Action, Field Equations and Solution}

\label{sec:2}

We consider EGB gravity in $D$-dimensions coupled to EH
NLED. The dynamics of the system is governed by the action \cite{Plebanski,
Salazar},  
\begin{equation}
\mathbf{S=}\frac{1}{16\pi }\int d^{D}x\sqrt{-g}\bigg[R-\frac{\left(
D-1\right) \left( D-2\right) }{3}\Lambda -4\mathcal{L}^{\mathrm{EH}%
}(F,G)+\alpha \mathcal{L}^{\mathrm{GB}}\bigg],  \label{action}
\end{equation}%
here we used cosmological constant ($\Lambda $), Ricci scalar ($R$), 
$g=\left( \det g_{\mu \nu }\right) $, and $\alpha $ is the coupling constant. The GB Lagrangian $%
\mathcal{L}^{\mathrm{GB}}$ is defined as
\begin{equation}
\mathcal{L}^{\mathrm{GB}}=R^{2}-4R_{ab}R^{ab}+R_{abcd}R^{abcd},
\end{equation}%
with Riemann tensor ($R_{abcd}$) and Ricci tensor ($R_{ab}$). The
EH Lagrangian $\mathcal{L}^{\mathrm{EH}}(F,G)$  takes the form 
\begin{equation}
\mathcal{L}^{\mathrm{EH}}\left( F,G\right) =-F+\frac{a}{2}F^{2}+\frac{7a}{8}%
G^{2},
\end{equation}%
the electromagnetic invariants are given by 
\begin{equation}
F=\frac{1}{4}F_{\mu \nu }F^{\mu \nu }\text{ \ and }G=\frac{1}{4}F_{\mu \nu
}^{\text{ \ \ }\ast }F^{\mu \nu },
\end{equation}
here $F_{\mu \nu }=\partial _{\mu }A_{\nu }-\partial _{\nu }A_{\mu }$ is
the usual Maxwell tensor. The parameter $a$ controls the strength of
EH NLED corrections. In order to simplify the NLED,
we employ the Legendre-dual formulation of NLED
introduced by Pleba\'{n}ski \cite{Plebanski}.  In this approach, one defines
the tensor $P_{\mu \nu }$
\begin{equation}
P_{\mu \nu }=\left( 1-aF\right) F_{\mu \nu }-^{\ast }F_{\mu \nu }\frac{7a}{4}%
G,
\end{equation}%
which allows one to construct two independent scalar invariants expressed as
\begin{equation}
s=-\frac{1}{4}P_{\mu \nu }P^{\mu \nu }\text{ \ \ and \ \ }t=-\frac{1}{4}%
P_{\mu \nu }^{\text{ \ \ }\ast }P^{\mu \nu }.
\end{equation}%
To first order in the parameter $a$, the Euler–Heisenberg Hamiltonian density takes the form
\begin{equation}
\mathcal{H}=s-\frac{a}{2}s^{2}-\frac{7a}{8}t^{2}.
\end{equation}%

The equations of the gravitational field are obtained by varying the action (\ref{action}) with respect to the metric.
\begin{equation}
R_{\mu \nu }-\frac{1}{2}g_{\mu \nu }R+\frac{\left( D-1\right) \left(
D-2\right) }{3}\Lambda +\alpha H_{\mu \nu }^{\mathrm{GB}}=T_{\mu \nu }^{%
\mathrm{EH}},  \label{Einstein}
\end{equation}%
supplemented by the electromagnetic field equation%
\begin{equation*}
\nabla _{\mu }P^{\mu \nu }=0.
\end{equation*}%
Here $H_{\mu \nu }^{\left( \mathrm{GB}\right) }$ is the Lanczos tensor \cite{Lanczos}
\begin{equation}
H_{\mu \nu }=2\left( -R_{\mu cd\lambda }R_{\nu }^{d\lambda
c}+RR_{\mu \nu }-2R_{\mu \nu bc}R^{dc}-2R_{\mu c}R_{\nu }^{c}\right) -\frac{1}{2}\mathcal{L%
}^{\mathrm{GB}}\text{ }g_{\mu \nu },
\end{equation}
while the stress-energy tensor of the EH field in the $P$
framework is given by \cite{Salazar}, 
\begin{equation}
T_{\mu \nu }^{\mathrm{EH}}=\left( 1-as\right) P_{\mu }^{\sigma }P_{\nu
\sigma }+g_{\mu \nu }\left( s-\frac{3a}{2}s^{2}-\frac{7a}{2}t^{2}\right),
\label{set}
\end{equation}
In the limit $a\rightarrow 0$, the standard Maxwell energy-momentum tensor
is recovered. Now, we consider the line element as
\begin{equation}
ds^{2}=-\mathcal{F}\left( r\right) dt^{2}+\frac{1}{\mathcal{F}\left(
r\right) }dr^{2}+r^{2}d\Omega _{D-2}^{2},
\end{equation}
where  $\left( D-2\right) $
dimensional unit sphere is denoted by $d\Omega _{D-2}^{2}$. It can be expressed as
\begin{equation}
d\Omega _{D-2}^{2}=d\theta _{D-3}^{2}+\sin \theta _{D-3}^{2}\left( d\theta
_{D-4}^{2}+\sin \theta _{D-4}^{2}\left( ...+\sin \theta _{2}^{2}\left(
d\theta _{1}^{2}+\sin \theta _{1}^{2}d\varphi \right) \right) \right) .
\end{equation}%
For this ansatz, the ($t$,$t$) component takes the form 
\begin{eqnarray}
T_{0}^{0} &=&\frac{D-2}{2}\left[ \frac{\mathcal{F}\left( r\right) ^{\prime }%
}{r}+\frac{\left( D-3\right) \mathcal{F}\left( r\right) }{r^{2}}-\frac{D-3}{%
r^{2}}+\frac{\left( D-1\right) }{3}\Lambda \right] -\frac{\alpha \left(
D-2\right) \left( D-3\right) \left( D-4\right) }{2}\times   \notag \\
&&\left[ \frac{2\mathcal{F}\left( r\right) \mathcal{F}\left( r\right)
^{\prime }}{r^{3}}-\frac{2\mathcal{F}\left( r\right) ^{\prime }}{r^{3}}+%
\frac{\left( D-5\right) \mathcal{F}\left( r\right) ^{2}}{r^{4}}-\frac{%
2\left( D-5\right) \mathcal{F}\left( r\right) }{r^{4}}+\frac{\left(
D-5\right) }{r^{4}}\right].  \label{EGBT}
\end{eqnarray}

It is interesting to mention that topologically GB term is referred as four dimensions, a nontrivial
contribution can be obtained by adopting the regularization procedure
proposed in \cite{Glavan}. Rescaling the coupling constant as $\alpha \rightarrow
\alpha /\left( D-4\right) $ and taking the limit $D\rightarrow 4$, one finds
\begin{equation}
T_{0}^{0}=\left[ \frac{\mathcal{F}\left( r\right) ^{\prime }}{r}+\frac{%
\mathcal{F}\left( r\right) }{r^{2}}-\frac{1}{r^{2}}\right] +\Lambda -\alpha %
\left[ \frac{2\mathcal{F}\left( r\right) \mathcal{F}\left( r\right) ^{\prime
}}{r^{3}}-\frac{2\mathcal{F}\left( r\right) ^{\prime }}{r^{3}}-\frac{%
\mathcal{F}\left( r\right) ^{2}}{r^{4}}+\frac{2\mathcal{F}\left( r\right) }{%
r^{4}}-\frac{1}{r^{4}}\right].
\end{equation}%
We restrict attention to a purely electric configuration with charge $Q$. In
this case, the only nonvanishing component of $P_{\mu \nu }$ is
\begin{equation}
P_{\mu \nu }=\frac{Q}{r^{2}}\left( \delta _{\mu }^{0}\delta _{\nu
}^{1}-\delta _{\mu }^{1}\delta _{\nu }^{0}\right) ,
\end{equation}%
which yields the electromagnetic invariants%
\begin{equation}
s=\frac{Q^{2}}{2r^{4}}\text{ \ and }t=0.
\end{equation}%
by using  $\mathcal{F}\left( r\right) =1+\mathcal{G}%
\left( r\right) $ the NLED reduce to 
\begin{equation}
\left( r^{3}-2\alpha r\mathcal{G}\left( r\right) \right) \mathcal{G}\left(
r\right) ^{\prime }+\left( r^{2}+\alpha \mathcal{G}\left( r\right) \right) 
\mathcal{G}\left( r\right) +\Lambda r^{4}+\frac{Q^{2}}{2}-\frac{a Q^{4}}{%
8r^{4}}=0.
\end{equation}%
Upon integration, one obtains the metric function describing a 4D EGB black
hole coupled to EH NLED
\begin{equation}
\mathcal{F}\left( r\right) =\frac{2\alpha+r^{2}}{2\alpha }\left( 1\pm \sqrt{1+%
\frac{4\alpha }{r^{2}}\left( \frac{2M}{r}-\frac{Q^{2}}{r^{2}}+\frac{\Lambda 
}{3}r^{2}+\frac{aQ^{4}}{20r^{6}}\right) }\right),  \label{fun}
\end{equation}%
The solution in Eq. (\ref{fun}) follows from the dimensional regularization scheme introduced in \cite%
{Glavan} and is compatible with the consistent Aoki-Gorji-Mukohyama theory
of 4D EGB gravity \cite{Zangeneh}. This class of BHs is parametrized
by the mass $M$, cosmological constant $\Lambda $, EH NLED
nonlinear parameter $a$, electric charge $Q$, and GB coupling constant $%
\alpha $. For $a=0$, Eq.(\ref{fun}) reduces to the BH solution
previously obtained in \cite{Pedro}. In order to analyze the general
structure of the solution (\ref{fun}), we consider the limit $\alpha
\rightarrow 0$. For the branch corresponding to the minus sign " $-$ ", the
metric function becomes 
\begin{equation}
\mathcal{F}\left( r\right) =1-\frac{2M}{r}+\frac{Q^{2}}{r^{2}}-\frac{\Lambda 
}{3}r^{2}-\frac{aQ^{4}}{20r^{6}},
\end{equation}%
which describes an asymptotically AdS BH coupled to
EH NLED. In contrast, for the branch
associated with the plus sign " $+$", we have
\begin{equation*}
\mathcal{F}\left( r\right) =1+\frac{2M}{r}-\frac{Q^{2}}{r^{2}}+\left( \frac{%
\Lambda }{3}+\frac{1}{\alpha }\right) r^{2}+\frac{aQ^{4}}{20r^{6}}.
\end{equation*}%

\begin{figure}[H]
\begin{minipage}[t]{0.5\textwidth}
        \centering
        \includegraphics[width=\textwidth]{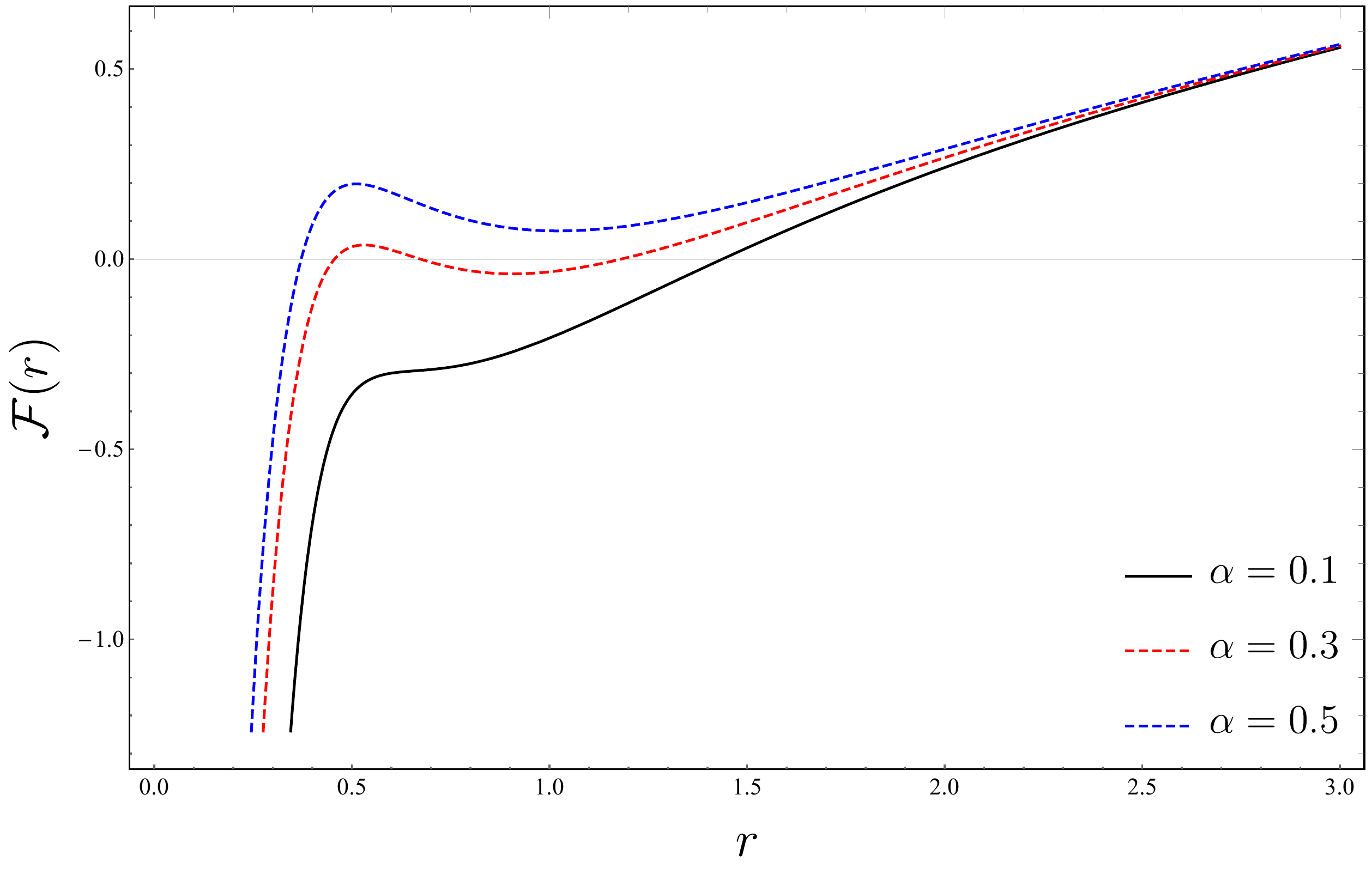}
                \subcaption{$a=0.5$ }
        \label{fig:met1}
\end{minipage}
\begin{minipage}[t]{0.5\textwidth}
        \centering
        \includegraphics[width=\textwidth]{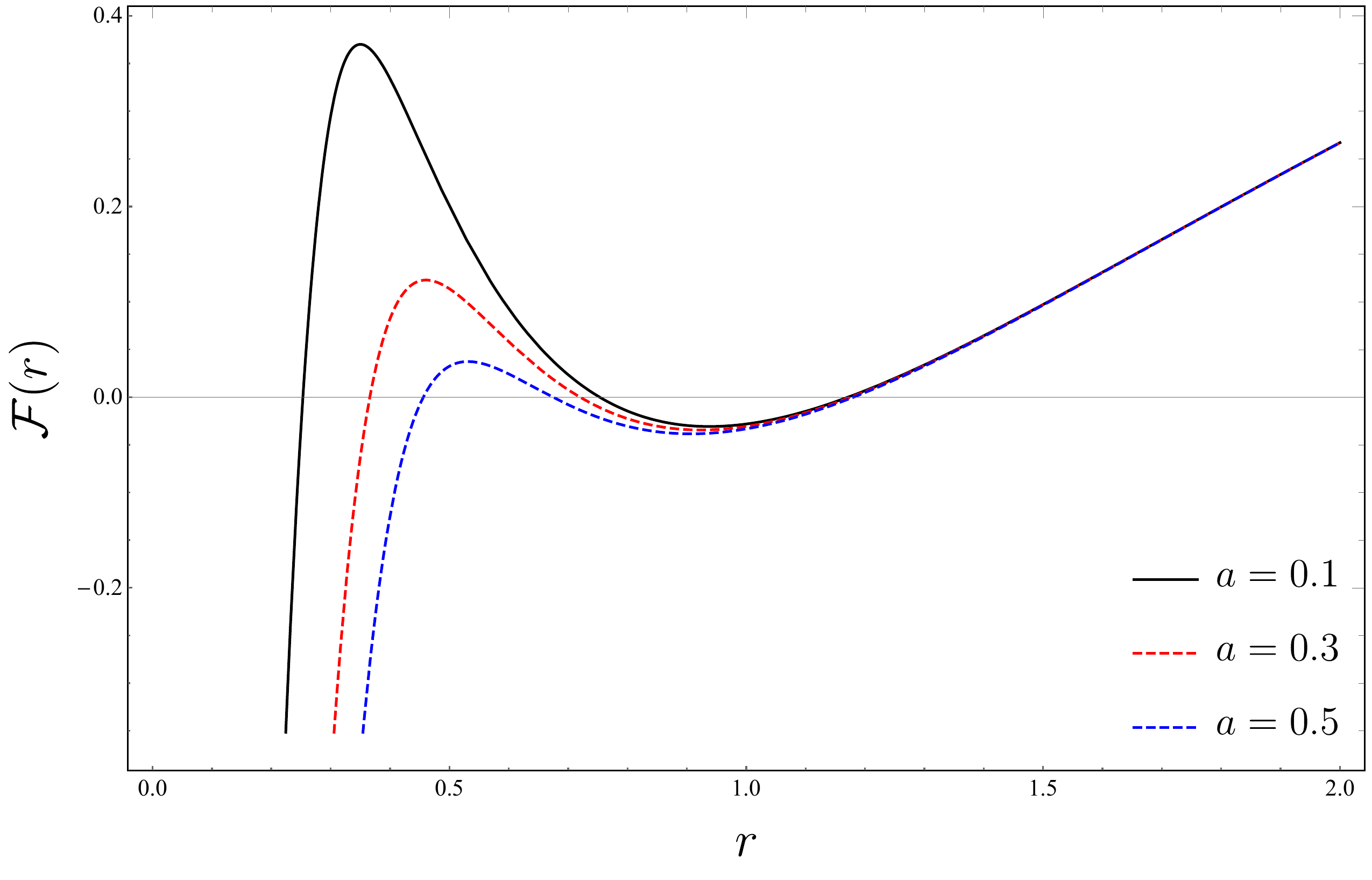}
                \subcaption{$\alpha=0.3$}
        \label{fig:met2}
\end{minipage}
\caption{Plots of $\mathcal{F}(r)$ versus $r$ for $M=1$ , $Q=0.9$ and $\Lambda=-0.05$ }
\label{fig:metric}
\end{figure}

This branch effectively reduces to a charged-(A)dS type
solution with negative gravitational mass and an imaginary electric charge,
rendering it physically unacceptable. Consequently, the minus branch
represents the physically meaningful solution. In what follows, we therefore
restrict our analysis to the negative branch only.
We now examine the horizon structure of the spacetime, which is determined
by the condition%
\begin{equation}
\mathcal{F}\left( r\right) =1+\frac{r^{2}}{2\alpha }\left( 1-\sqrt{1+\frac{%
4\alpha }{r^{2}}\left( \frac{2M}{r}-\frac{Q^{2}}{r^{2}}+\frac{\Lambda }{3}%
r^{2}+\frac{aQ^{4}}{20r^{6}}\right) }\right) =0.
\end{equation}%
The roots of this equation determine the horizon structure of the BH.
In general, the BH admits three positive real roots: the event
horizon, $r_{+}$, the inner (Cauchy) horizon, $r_{-}$, and the cosmological
horizon, $r_{C}$. Figure \ref{fig:metric} displays the behavior of the metric function $%
\mathcal{F}\left( r\right) $ for different values of the parameters $\alpha $%
 and $a$, which encode the effects of EGB gravity and NLED,
respectively. 
\section{Thermodynamics}
\label{sec:3}
\begin{figure}[H]
\begin{minipage}[t]{0.5\textwidth}
        \centering
        \includegraphics[width=\textwidth]{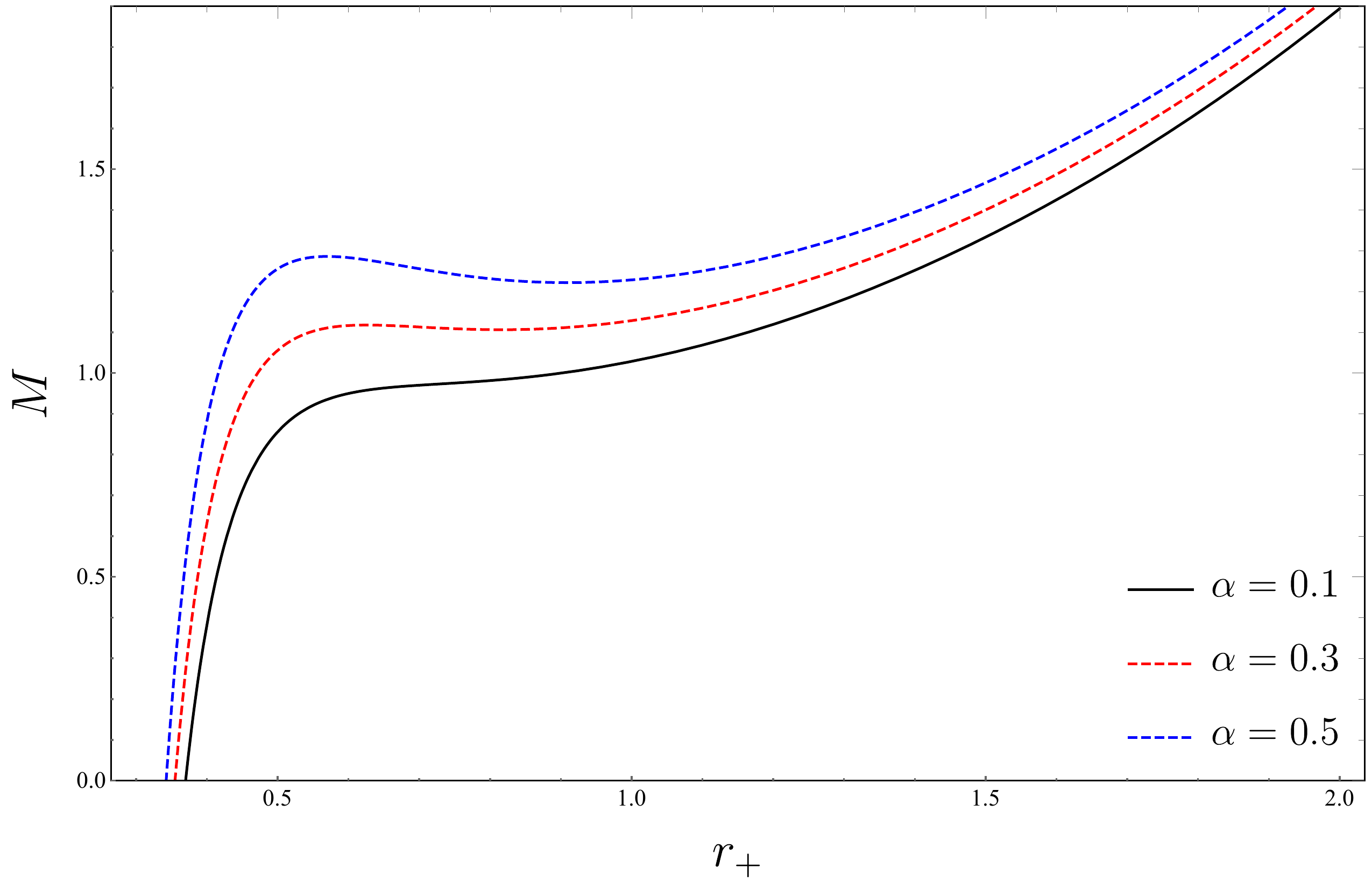}
                \subcaption{$a=0.6$ }
        \label{fig:m1}
\end{minipage}
\begin{minipage}[t]{0.5\textwidth}
        \centering
        \includegraphics[width=\textwidth]{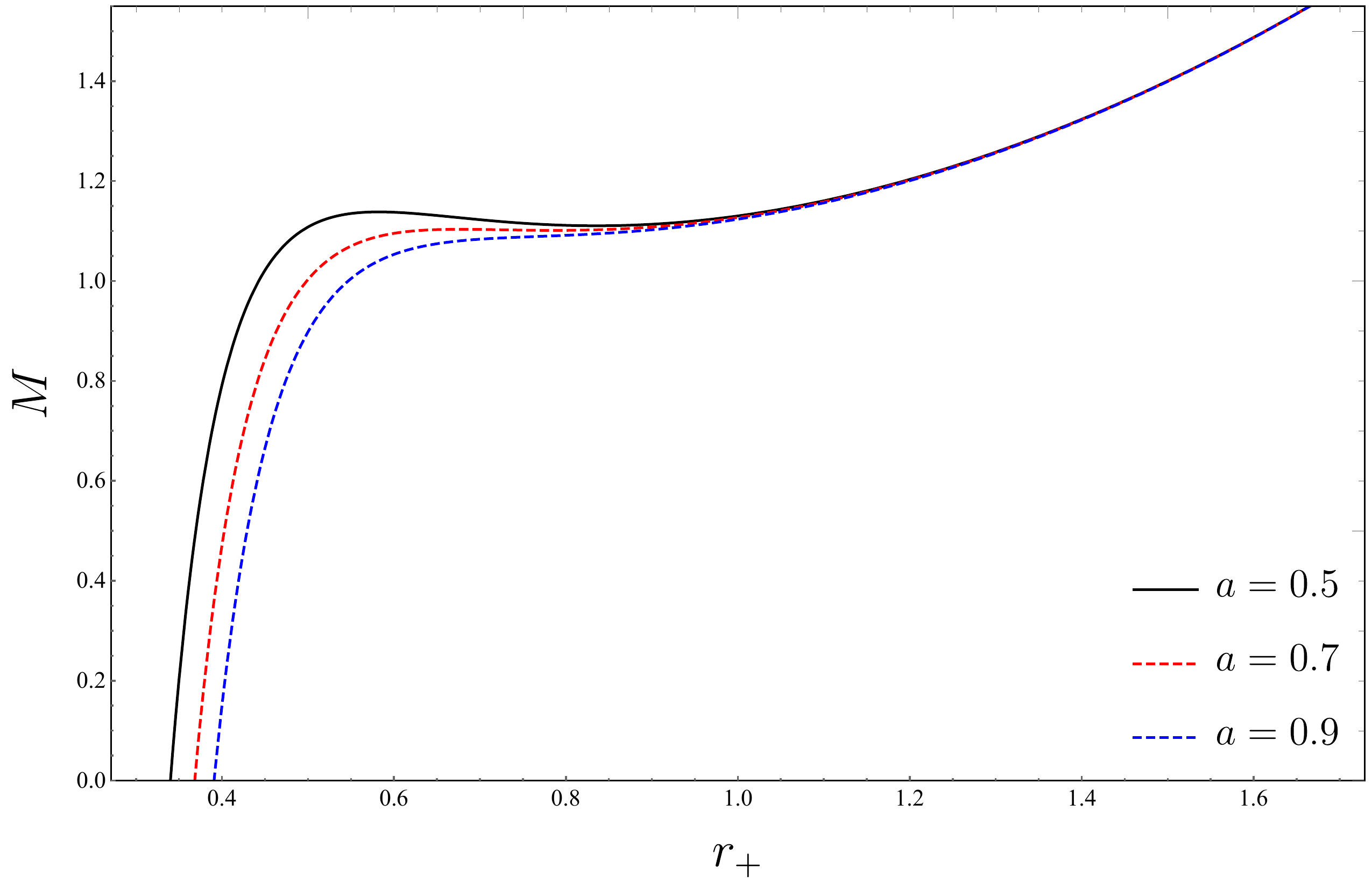}
                \subcaption{$\alpha=0.3$}
        \label{fig:m2}
\end{minipage}
\caption{Plots of $M$ versus $r_+$ for  $Q=0.9$ and $\Lambda=-0.5$ }
\label{fig:mass}
\end{figure}

In this section, we investigate the
thermodynamic properties of EH AdS BH in 4D EGB
gravity. The pioneering works of Bekenstein and Hawking \cite{HawkingSW,JacobD} established a
fundamental connection between thermodynamic quantities and the geometric
properties of BHs, identifying the entropy with the area of the
event horizon and the temperature with the surface gravity.
Consequently, all thermodynamic quantities of the BH are evaluated
at the event horizon $r_{+}$. In particular, the BH mass can be
obtained from the horizon condition $\mathcal{F}\left( r_{+}\right) =0$,
yielding 
\begin{equation}
M=\frac{r_{+}}{2}\left( 1+\frac{\alpha +Q^{2}}{r_{+}^{2}}-\frac{\Lambda
r_{+}^{2}}{3}-\frac{aQ^{4}}{20r_{+}^{6}}\right).
\end{equation}%
Figure \ref{fig:mass} displays the variation of the BH mass as a function of the
event horizon radius $r_{+}$. The mass exhibits a non-monotonic dependence
on $r_{+}$, characterized by the presence of a local maximum and a local
minimum. These extrema correspond to turning points separating different
BH branches. We find that increasing the GB coupling $\alpha $
enhances the maximum value of the mass and shifts its location toward
smaller horizon radii, indicating that higher-curvature effects become
dominant at shorter length scales. In contrast, increasing the
EH parameter also increase the maximum mass but shift its
location toward larger horizon radii. This behavior reflects the role of
nonlinear field effects in delaying the onset of thermodynamic instability
and extending the region of stability for small BHs. Consequently,
the competing effects of higher-curvature gravity and nonlinear
electrodynamics give rise to a rich phase structure and significantly modify
the thermodynamic stability of the system.
Further, we can define the Hawking temperature ($T_{H}$) in terms of the BH can be surface gravity ($\kappa $) as 
\begin{equation}
T_{H}=\frac{\kappa }{2\pi }=\frac{1}{4\pi }\left. \frac{\partial }{\partial r%
}\mathcal{F}\left( r\right) \right\vert _{r=r_{+}}=\frac{1-\frac{\alpha
+Q^{2}}{r_{+}^{2}}-\Lambda r_{+}^{2}+\frac{aQ^{4}}{4r_{+}^{6}}}{4\pi
r_{+}\left( 1+\frac{2\alpha }{r_{+}^{2}}\right) }.  \label{TT}
\end{equation}%
Through the bounds of extended phase space, the negative value of $\Lambda $ can be interpreted as a thermodynamic pressure $ P $, defined as
\begin{equation}
P=-\frac{\Lambda }{8\pi }.  \label{PP}
\end{equation}
Substituting (\ref{PP}) into Eq. (\ref{TT}), the Hawking temperature can be
rewritten as%
\begin{equation}
T_{H}=\frac{1-\frac{\alpha +Q^{2}}{r_{+}^{2}}+8\pi Pr_{+}^{2}+\frac{aQ^{4}}{%
4r_{+}^{6}}}{4\pi r_{+}\left( 1+\frac{2\alpha }{r_{+}^{2}}\right) }.\label{th}
\end{equation}

\begin{figure}[H]
\begin{minipage}[t]{0.5\textwidth}
        \centering
        \includegraphics[width=\textwidth]{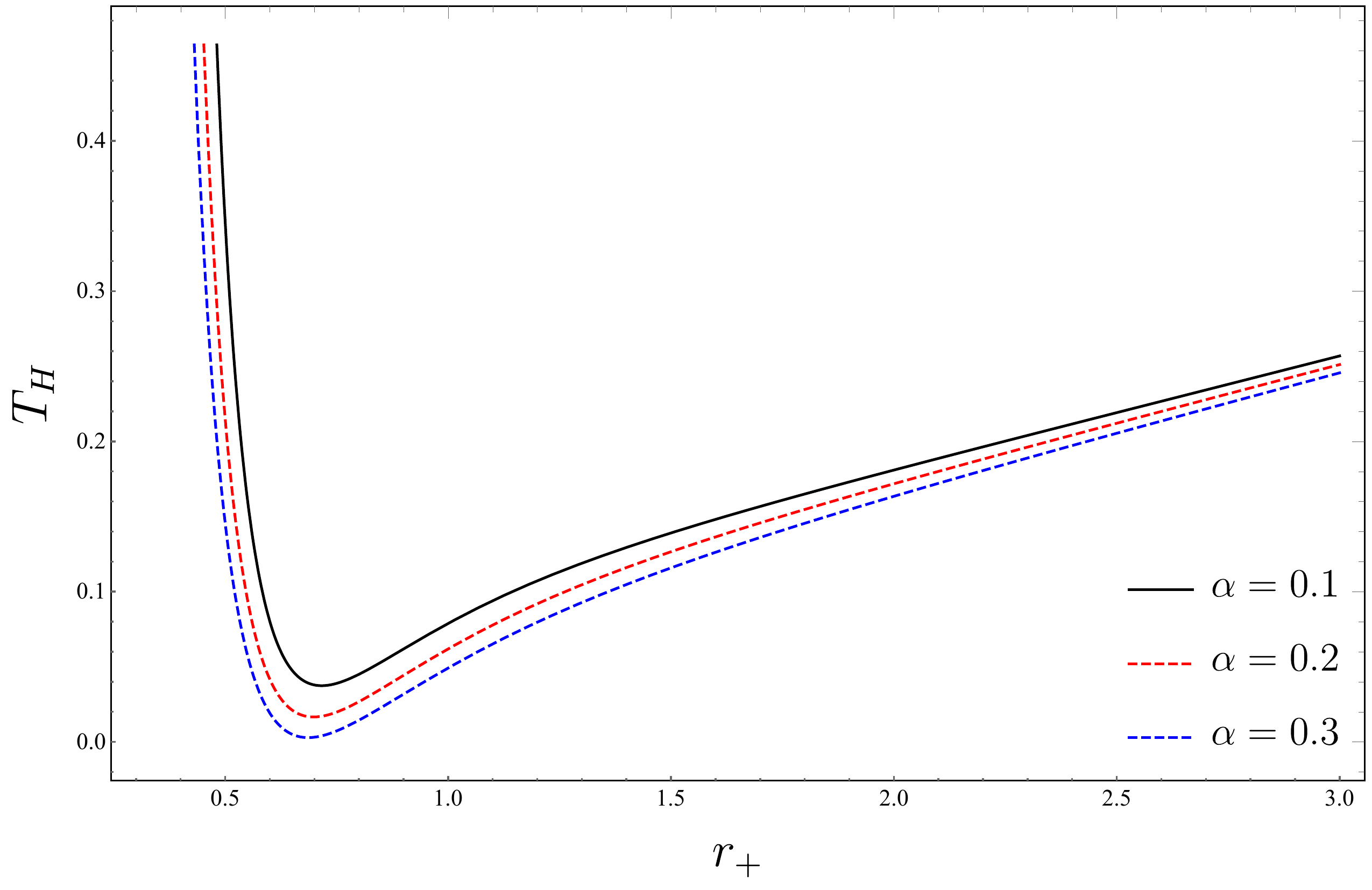}
                \subcaption{$a=0.6$ }
        \label{fig:t1}
\end{minipage}
\begin{minipage}[t]{0.5\textwidth}
        \centering
        \includegraphics[width=\textwidth]{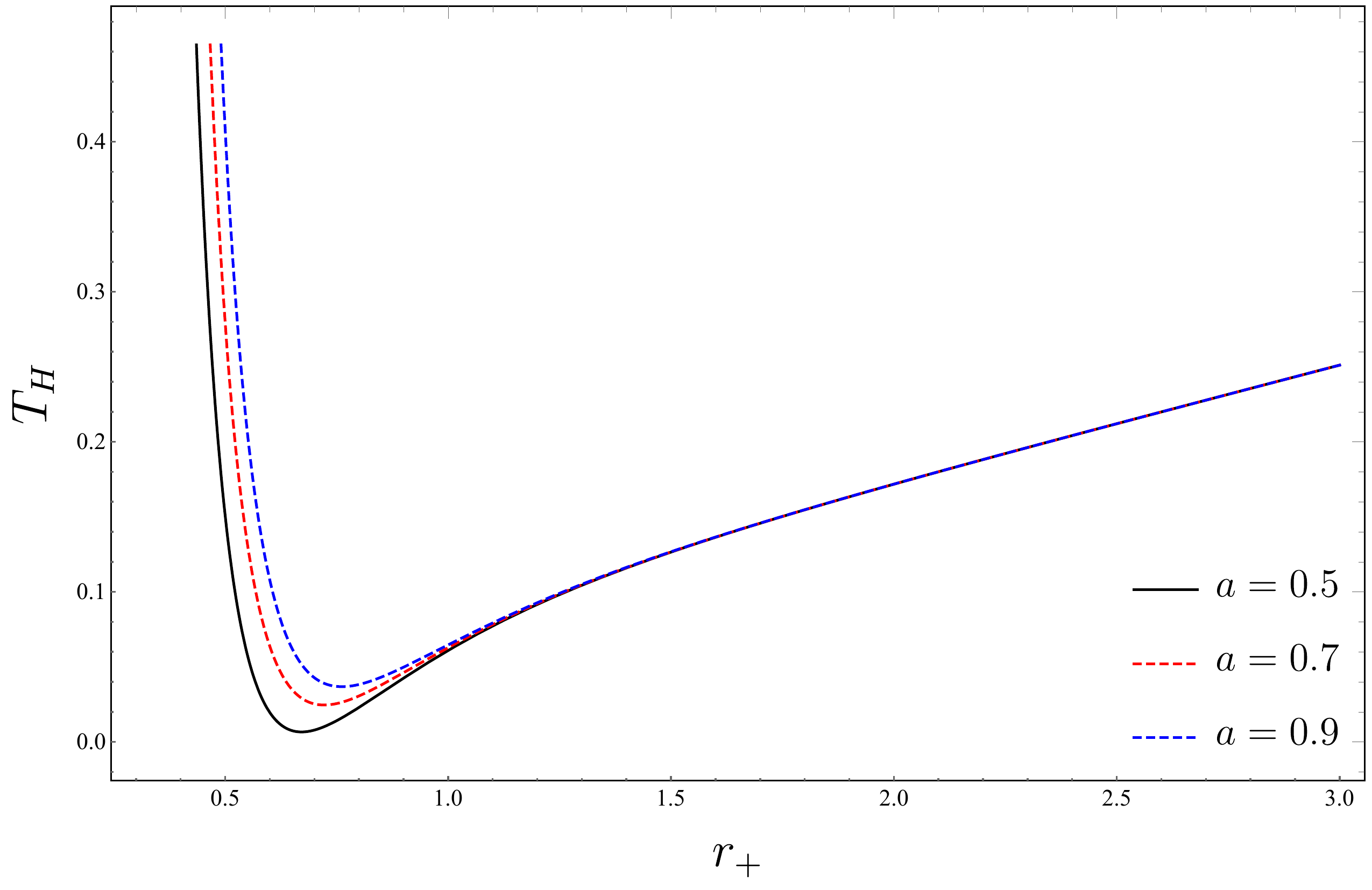}
                \subcaption{$\alpha=0.2$}
        \label{fig:t2}
\end{minipage}
\caption{Plots of $T_H$ versus $r_{+}$ for $Q=0.9$ and $P=\frac{1}{8\pi}$ }
\label{fig:tem}
\end{figure}

\begin{figure}[H]
\begin{minipage}[t]{0.5\textwidth}
        \centering
        \includegraphics[width=\textwidth]{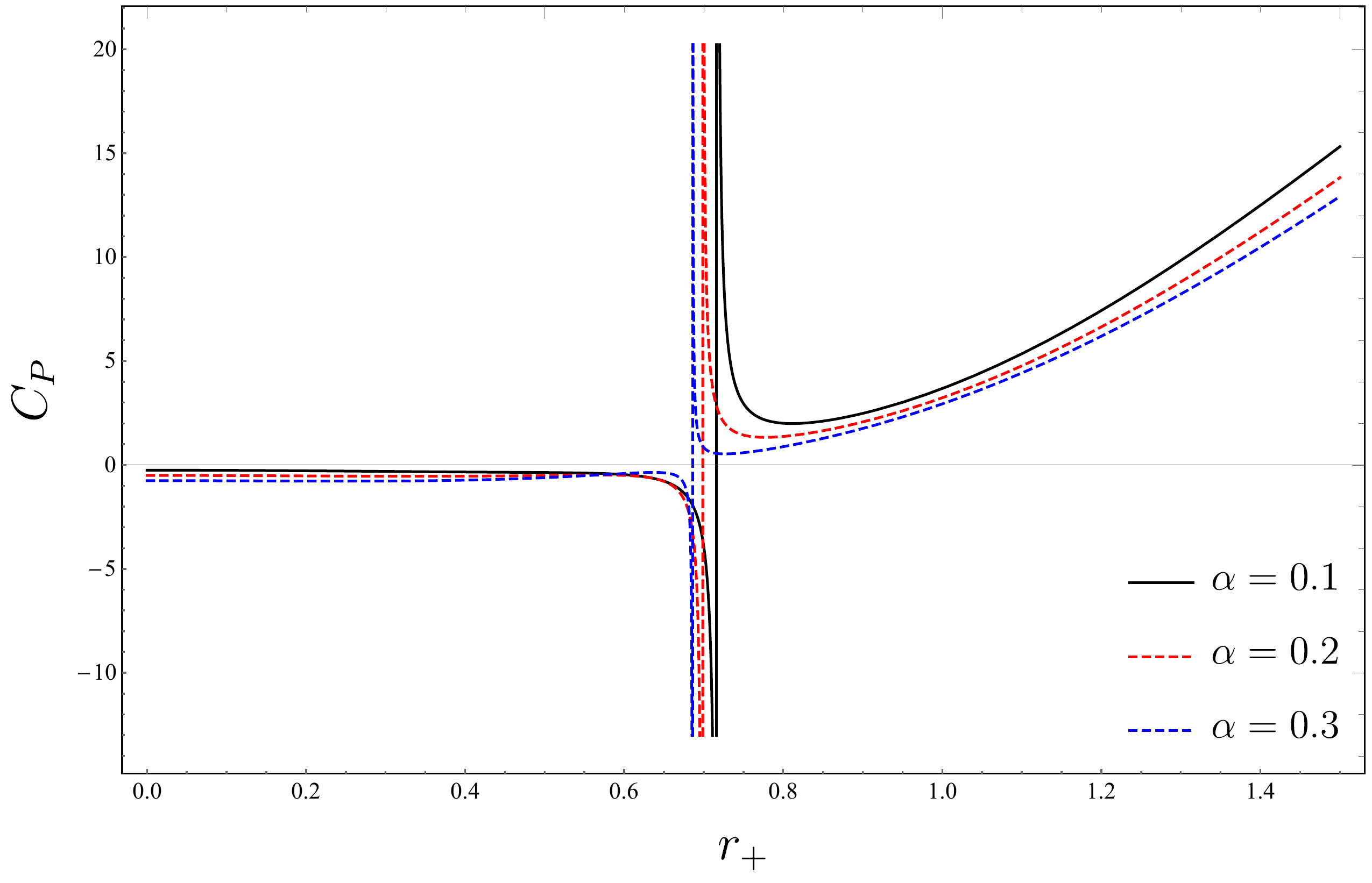}
                \subcaption{$a=0.6$ }
        \label{fig:h1}
\end{minipage}
\begin{minipage}[t]{0.5\textwidth}
        \centering
        \includegraphics[width=\textwidth]{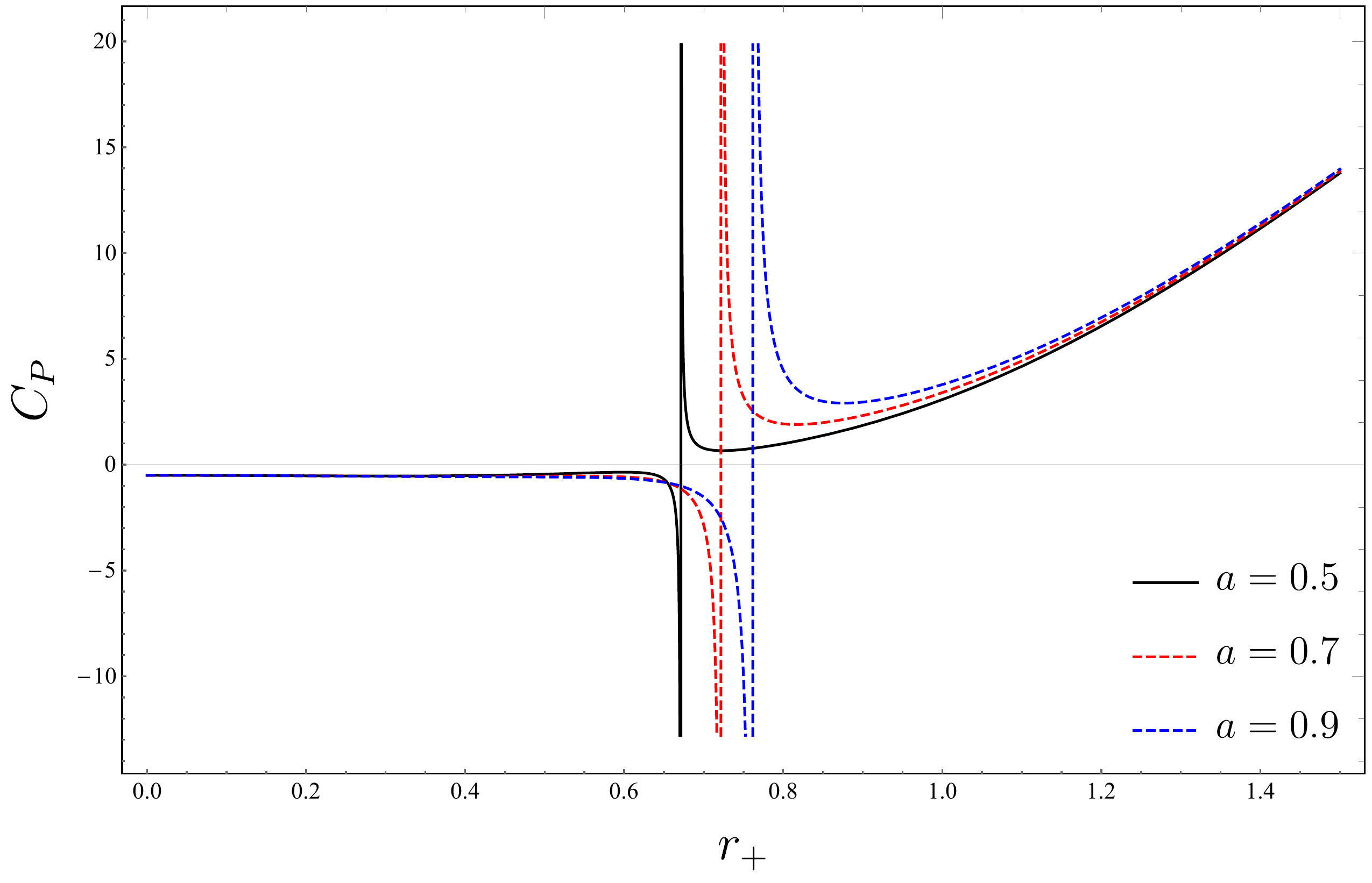}
                \subcaption{$\alpha=0.2$}
        \label{fig:h2}
\end{minipage}
\caption{Plots of $C_P$ versus $r_{+}$ for $Q=0.9$ and $P=\frac{1}{8\pi}$ }
\label{fig:heat}
\end{figure}

\begin{figure}[H]
\begin{minipage}[t]{0.5\textwidth}
        \centering
        \includegraphics[width=\textwidth]{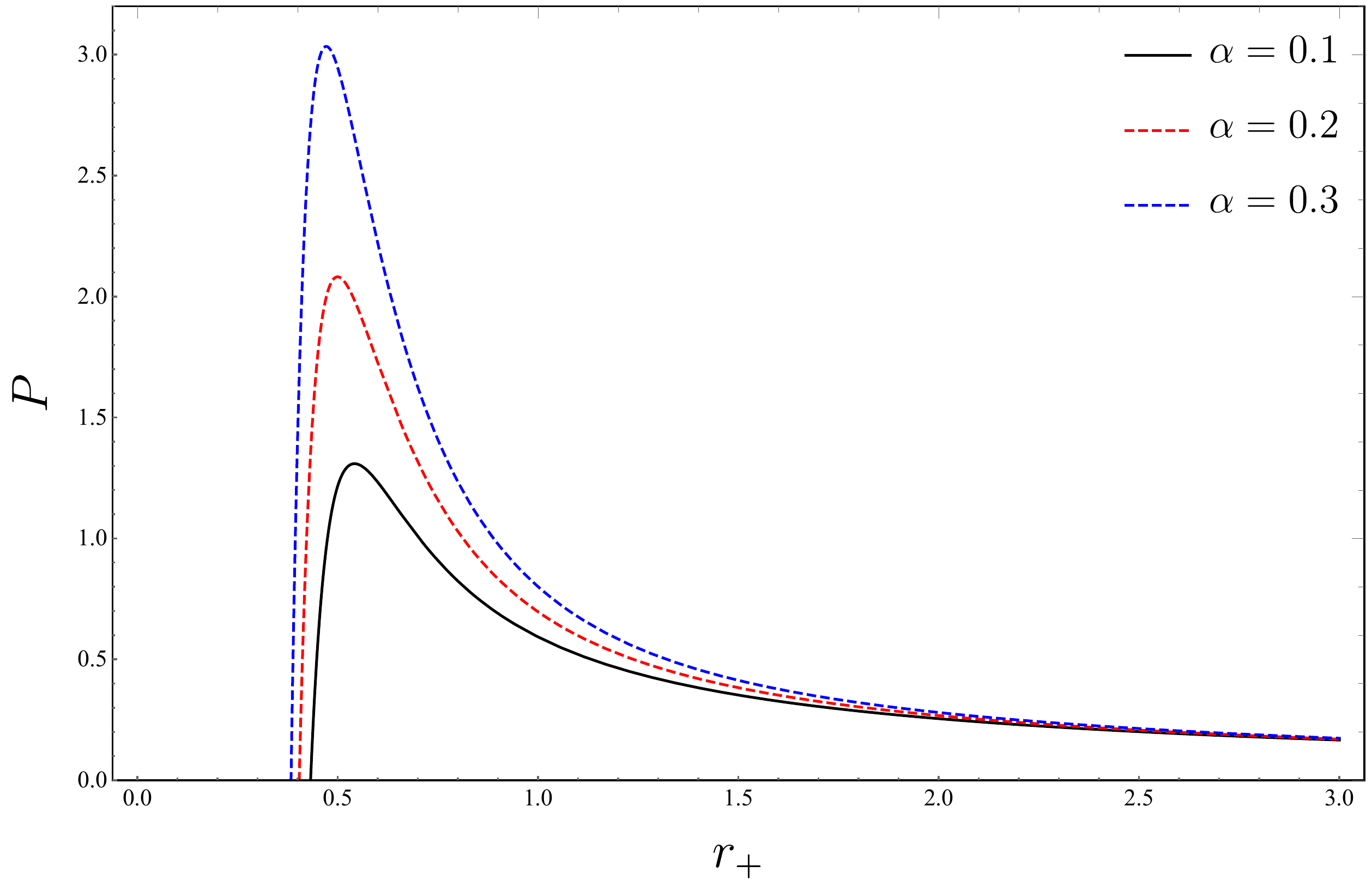}
                \subcaption{$a=0.6$ }
        \label{fig:p1}
\end{minipage}
\begin{minipage}[t]{0.5\textwidth}
        \centering
        \includegraphics[width=\textwidth]{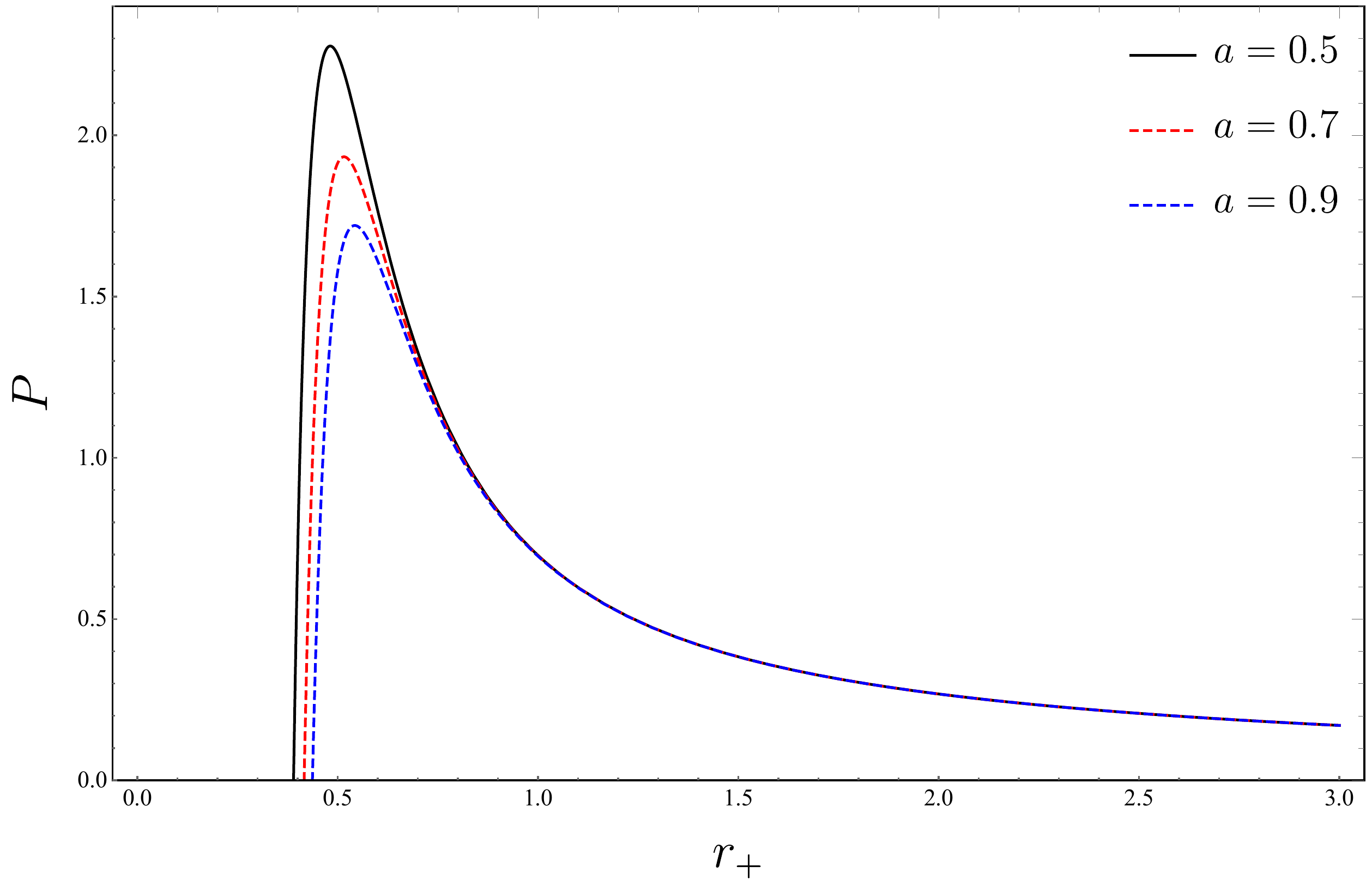}
                \subcaption{$\alpha=0.2$}
        \label{fig:p2}
\end{minipage}
\caption{Pressure $P$ versus horizon radius $r_{+}$ for $Q=0.9$ and $P=\frac{1}{8\pi}$ }
\label{fig:pert}
\end{figure}

In the absence of EGB gravity ($\alpha =0$), we can obtain
\begin{equation}
T_{H}=\frac{1}{4\pi r_{+}}\left( 1-\frac{Q^{2}}{r_{+}^{2}}+8\pi Pr_{+}^{2}+%
\frac{aQ^{4}}{4r_{+}^{6}}\right),
\end{equation}
which agrees with the result obtained in \cite{Magos}. Furthermore, by setting $a=0$,
one recovers the standard Hawking temperature of the charged AdS BH
\begin{equation}
T_{H}=\frac{1}{4\pi r_{+}}\left( 1-\frac{Q^{2}}{r_{+}^{2}}+8\pi
Pr_{+}^{2}\right) .
\end{equation}%
Clearly, $T_{H}$ of the EH AdS BH in
4D EGB gravity is modified by both the GB and EH parameters,
indicating that the temperature is highly sensitive to higher-curvature and
NLED effects (see Fig. \ref{fig:tem}). Figure \ref{fig:tem} displays the Hawking temperature as a
function of the horizon radius $r_{+}$. The temperature exhibits a minimum, separating small and large BH phases. Decreasing the GB coupling $\alpha$ raises the temperature and shifts the minimum toward smaller horizon radii, indicating the dominance of higher-curvature effects at short length scales. In contrast, increasing the EH parameter shifts the minimum toward larger $r_{+}$ reflecting the influence of NLED. These results show that higher-curvature and NLED corrections affect the thermal behavior in competing ways.
\vspace{0.1cm}
In the framework of  extended phase
space, the generalized first law of BH for the considered BH can be expressed as
\begin{equation}
dM=TdS+\Phi dQ+VdP.  \label{25}
\end{equation}%
Using the first law in Eq. (\ref{25}), the Bekenstein entropy is obtained as%
\begin{equation}
S=\int \frac{dM}{T}=\int 2\pi r_{+}\left( 1+\frac{2\alpha }{r_{+}^{2}}%
\right) dr_{+}=\pi r_{+}^{2}+4\alpha \ln r_{+}.
\end{equation}
The logarithmic correction encodes the contribution of higher-order curvature terms associated with the GB invariant, resulting in a modification of the standard area law.
Another notable example of a quantity in thermodynamics is the specific heat capacity at constant pressure $C_{P}$, which describes the local thermodynamic stability of the BH. In the present case, it is given by
\begin{equation}
C_{P}=\left( \frac{dM}{dT}\right) _{P}=-\frac{2\pi r_{+}^{2}\left( 1+\frac{%
2\alpha }{r_{+}^{2}}\right) ^{2}\left( 1+\frac{aQ^{4}}{4r_{+}^{6}}-\frac{%
Q^{2}+\alpha }{r_{+}^{2}}+8\pi Pr_{+}^{2}\right) }{\left( 1-3\frac{\alpha
+Q^{2}}{r_{+}^{2}}-8\pi Pr_{+}^{2}+\frac{7aQ^{4}}{4r_{+}^{6}}\right) -\frac{%
2\alpha }{r^{2}}\left( 1+\frac{Q^{2}+\alpha }{r_{+}^{2}}+24\pi Pr_{+}^{2}-%
\frac{5aQ^{4}}{4r_{+}^{6}}\right) }.
\end{equation}%
The local thermodynamic stability is governed by the sign of the heat capacity, whose divergence occurs when $\frac{dT}{dr_{+}}=0$ corresponding to the minimum of the Hawking temperature at  $r_{+}=r_{\min }$. For $r_{+}<r_{\min }$ the heat capacity is negative (Fig.\ref{fig:heat}), indicating thermodynamic instability, while for $r_{+}>r_{\min }$ it becomes positive, signaling a stable large BH phase; this divergence therefore marks a second-order phase transition. Increasing the GB coupling shifts the divergence point toward smaller horizon radii, reflecting the dominance of higher-curvature effects at short length scales. In contrast, increasing the EH parameter moves the divergence toward larger $r_{+}$, demonstrating the competing influence of NLED on the thermodynamic phase structure.
Now, using Eq. (\ref{th}), we derive the equation of state, which after
rearrangement takes the form%
\begin{equation}
P=\frac{1}{2r_{+}}\left( 1+\frac{2\alpha }{r_{+}^{2}}\right) T_{H}+\frac{%
\alpha +Q^{2}}{8\pi r_{+}^{4}}-\frac{aQ^{4}}{32\pi r_{+}^{8}}-\frac{1}{8\pi
r_{+}^{2}}.  \label{p23}
\end{equation}%
Therefore, Eq.(\ref{p23}) can be regarded as $P=P\left( r_{+},T\right) $,
representing the equation of state in the extended phase space. This
expression is usually referred to as the geometric equation of state. Fig.\ref{fig:pert} displays the pressure isotherms in the extended phase space, revealing a non-monotonic behavior characteristic of critical phenomena. The maximum of each isotherm corresponds to the critical point separating distinct thermodynamic phases. Increasing the GB coupling raises the peak pressure, indicating that higher-curvature effects enhance the effective repulsive interactions and shift the critical point to higher pressures. In contrast, increasing the Euler–Heisenberg parameter lowers the maximum pressure, reflecting the screening effect of NLED and shifting the critical point toward lower pressures.

The critical parameters can be obtained by requiring
\begin{equation}
\left. \frac{\partial }{\partial r}P\right\vert _{r=r_{c}}=0,\left. \frac{%
\partial ^{2}}{\partial r^{2}}P\right\vert _{r=r_{c}}=0.
\end{equation}%
For these conditions, we can obtain the critical values as
\begin{equation}
T_{c}=\frac{1+\frac{aQ^{4}}{r_{c}^{6}}-2\frac{Q^{2}+\alpha }{r_{c}^{2}}}{%
2\pi r_{c}\left( 1+\frac{6\alpha }{r_{c}^{2}}\right) },
\end{equation}%
\begin{equation}
P_{c}=-\frac{aQ^{4}}{32\pi r_{c}^{8}}+\frac{\left( 1+\frac{2\alpha }{%
r_{c}^{2}}\right) \left( 1+\frac{aQ^{4}}{r_{c}^{6}}-2\frac{\left(
Q^{2}+\alpha \right) }{r_{c}^{2}}\right) }{4\pi r_{c}^{2}\left( 1+\frac{%
6\alpha }{r_{c}^{2}}\right) }+\frac{\alpha +Q^{2}}{8\pi r_{c}^{4}}-\frac{1}{%
8\pi r_{c}^{2}}.
\end{equation}
Here $r_{c}$ is the solution of the following algebraic equation: 
\begin{equation}
12\alpha ^{2}r_{c}^{4}+6Q^{2}r_{c}^{6}\left( 1+\frac{2\alpha }{r_{c}^{2}}%
\right) -r_{c}^{8}+12\alpha r_{c}^{6}-aQ^{4}\left(30\alpha
+7r_{c}^{2}\right)=0.
\end{equation}%
Since an analytic solution of this equation is not tractable, we solve it
numerically. The resulting critical parameters for varying $\alpha $ and EH couplings are summarized in Table 
\ref{tab:rtpc}. As shown in the table, increasing $\alpha $, leads to a monotonic increase in the critical
radius $r_{c}$, while both $T_{c}$ and $P_{c}$ decrease. This behavior indicates that the GB term significantly
affects the location of the critical point. In contrast, varying the
EH parameter $a$ produces only small changes in the critical
quantities. In particular, $r_{c}$ decreases slightly, whereas $T_{c}$ and $%
P_{c}$ increase marginally as $a$ increases, suggesting that the influence
of the EH correction on the critical behavior is comparatively
weak.

\begin{table}[H]
\centering%
\begin{tabular}{|c|c|c|c||c|c|c|c|}
\hline
\multicolumn{4}{|c||}{$a =0.4$} & \multicolumn{4}{c}{$\alpha=0.3$} \\ \hline
$\alpha $ & $r_{c}$ & $T_{c}$ & $P_{c}$ & $a$ & $r_{c}$ & $T_{c}$ & $P_{c}$
\\ \hline
0.1 & 2.27274 & 4.4838$\times 10^{-2}$ & 2.80444$\times 10^{-3}$ & 0.1 & 
2.80444 & 3.51464$\times 10^{-2}$ & 2.28976$\times 10^{-3}$ \\ 
0.3 & 2.80169 & 3.51581$\times 10^{-2}$ & 2.29168$\times 10^{-3}$ & 0.3 & 
2.80261 & 3.51542$\times 10^{-2}$ & 2.29104$\times 10^{-3}$ \\ 
0.5 & 3.2373 & 2.99118$\times 10^{-2}$ & 1.67694$\times 10^{-3}$ & 0.5 & 
2.80076 & 3.5162$\times 10^{-2}$ & 2.29233$\times 10^{-3}$ \\ 
0.7 & 3.61823 & 2.64878$\times 10^{-2}$ & 1.32353$\times 10^{-3}$ & 0.7 & 
2.79891 & 3.51699$\times 10^{-2}$ & 2.29362$\times 10^{-3}$ \\ 
0.9 & 3.96156 & 2.40254$\times 10^{-2}$ & 1.09356$\times 10^{-3}$ & 0.9 & 
2.79704 & 3.51778$\times 10^{-2}$ & 2.29493$\times 10^{-3}$ \\ \hline
\end{tabular}%
\caption{Critical quantities $r_{c}$, $T_{c}$ and $P_{c}$ for varying $\alpha $ and the EH parameter.}
\label{tab:rtpc}
\end{table}

\section{Joule Thomson Expansion}
\label{sec:4}
In this section, we analyze the Joule Thomson (JT) expansion in the scenario of considered BH. The JT mechanism constitutes a key thermodynamic pathway in which enthalpy remains invariant while temperature adjusts according to pressure variations. Within the BH formalism, it corresponds to an isenthalpic transformation in the extended phase space. The governing measure is the JT coefficient $\mu_{J}$, which relates pressure changes with temperature evolution and denotes whether the BH undergoes thermal contraction or expansion. Making use of the definitions of thermodynamic pressure, entropy and temperature, the BH EoS can be written as
\begin{eqnarray}\label{3-1}
\mu_{J}={\Big(\frac{\partial T}{\partial P}\Big)}_{H},
\end{eqnarray}
where $H$ denotes the enthalpy. When the expansion causes a decrease in pressure, the temperature may either rise or drop, with the sign of $\mu$ specifying the outcome: $\mu<0$ corresponds to heating, while $\mu>0$ signals cooling. Following the formulation, the JT coefficient can be recast in the alternative form
\begin{equation}\label{3-2}
\mu_{J}={\Big(\frac{\partial T}{\partial
P}\Big)}_{M}=\frac{1}{C_P}\Big[T\Big({\frac{\partial V}{\partial
T}}\Big)_P-V\Big],
\end{equation}
where $C_P$ is the specific heat at constant pressure. Also, the condition for obtaining the inversion temperature arises from setting $\mu=0$, yielding
\begin{equation}\label{3-3}
T_{i}=V {\Big(\frac{\partial T}{\partial V}\Big)}_{P}.
\end{equation}

\begin{figure}[H]\centering
\includegraphics[width=.46\linewidth,height=2.6in]{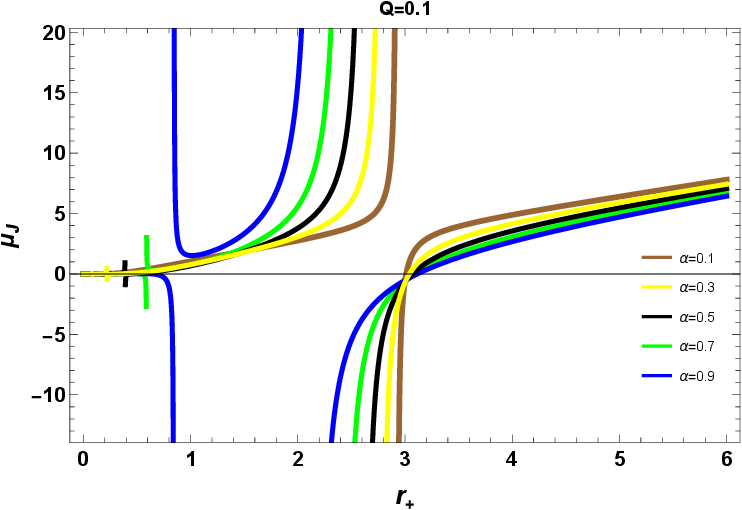}
\includegraphics[width=.46\linewidth,height=2.6in]{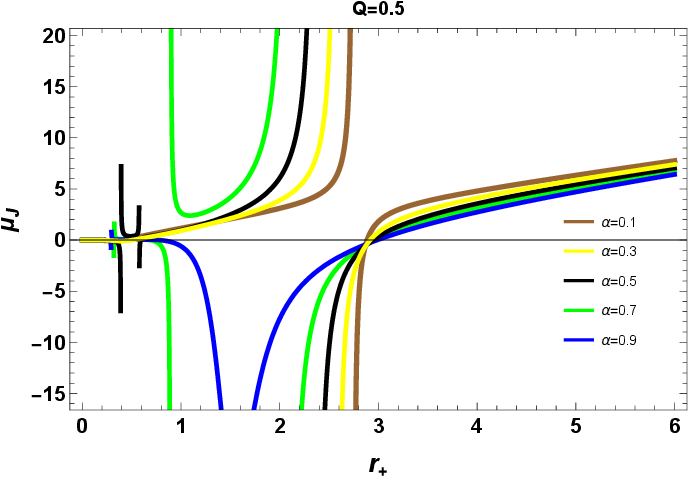}
\caption{\label{fa}Plots of $\mu_J$ versus $r_+$ with $a=0.9,M=1$.}
\label{fig:per}
\end{figure}

In this section, we investigate the JT expansion of the considered BH within the extended phase space framework that incorporates NLED effects. In such a thermodynamic setting, the variation of temperature with respect to pressure defines the expansion process, whereas the enthalpy remains unchanged. Since in AdS backgrounds the BH mass is interpreted as enthalpy, it consequently remains fixed during the expansion process. A straightforward approach to evaluate the JT coefficient $\mu$ can be achieved by carefully analyzing the relations among the thermodynamic quantities. In this case, the pressure can be recast in terms of the event horizon radius, the coupling constant and the mass of the BH, illustrated as
\begin{equation}\label{3-2-1}
P({M},r_+)=\frac{60 r^4_+ \left(\alpha -2 M r_++Q^2+r_+^2\right)-3 a Q^4}{160 \pi  r_+^8},
\end{equation}
and substituting this form of $P(M,r_+)$ into the temperature relation (\ref{th}), the expression for the Hawking temperature becomes
\begin{equation}\label{3-2-2}
T({M},r_+)=\frac{a Q^4-5 r_+^4 \left(2 \alpha -3 M r+2 Q^2+r_+^2\right)}{10 r_+^5 \left(2 \alpha +\pi  r_+^2\right)}.
\end{equation}
By employing the relations that connect the pressure and temperature, one can evaluate the JT coefficient $\mu_J$, which is defined as
\begin{equation}\label{3-2-3}
\mu_J={\Big(\frac{\partial T}{\partial
P}\Big)}_{M}={\Big(\frac{\partial T}{\partial
{r_{+}}}\Big)}_{M}{\Big(\frac{\partial {r_{+}}}{\partial
P}\Big)}_{M}=\frac{\Big(\partial T/\partial
{r_{+}\Big)}_{M}}{\Big(\partial P/\partial {r_{+}}\Big)_{M}}.
\end{equation}
This expression reduces to
\begin{eqnarray}\label{3-2-3s}
\mu_J=\frac{4 \pi  r_+^3 \left(a Q^4 \left(8 \alpha +5 \pi  r_+^2\right)-4 r_+^4 \left(4 \alpha ^2+\pi  r_+^2 \left(8 \pi  P r_+^4+3 Q^2-2 r_+^2\right)+4 \alpha  Q^2+(3 \pi -2) \alpha  r_+^2\right)\right)}{3 \left(2 \alpha +\pi  r_+^2\right)^2 \left(4 r_+^4 \left(\alpha +8 \pi  P r_+^4+Q^2-r_+^2\right)-a Q^4\right)}.
\end{eqnarray}

Figure~(\ref{fa}) illustrates how the JT coefficient, \(\mu_J\), changes depending on the event horizon radius \(r_+\) for various values of GB parameter \(\alpha\) and electric charge \(Q\). It can be noted that the greater the \(\alpha\), the more the inversion points move towards greater horizon radii which means the corrections with respect to the higher curvature are more dominant on the expansion factor. The charge \(Q\) also greatly influences the \(\mu_J\) values, which supports more of the electro-magnetic component of the response to the changes in the thermodynamic system. The change of sign in \(\mu_J\) distinguishes the cooling and heating sections and it indicates that the JT expansion is not trivial. This illustrates the combined influences of curvature and charge on the thermodynamic properties of BHs.

\begin{figure}[H]\centering
\includegraphics[width=.46\linewidth,height=2.6in]{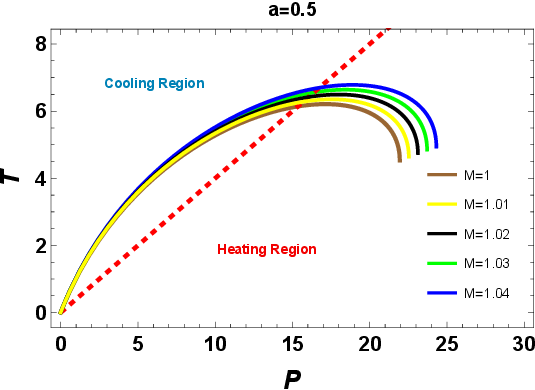}
\includegraphics[width=.46\linewidth,height=2.6in]{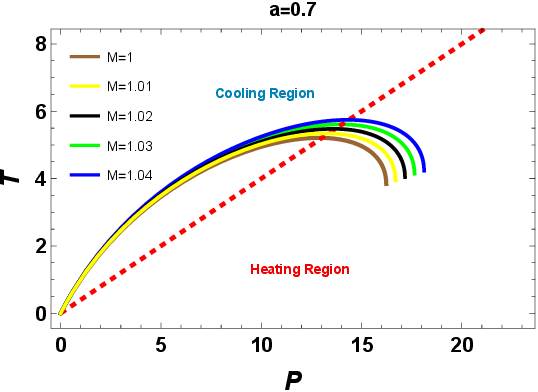}
\includegraphics[width=.46\linewidth,height=2.6in]{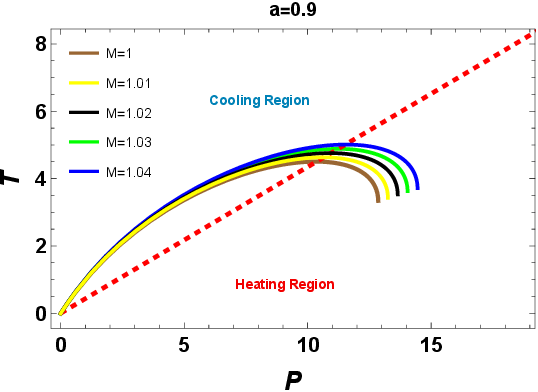}
\includegraphics[width=.46\linewidth,height=2.6in]{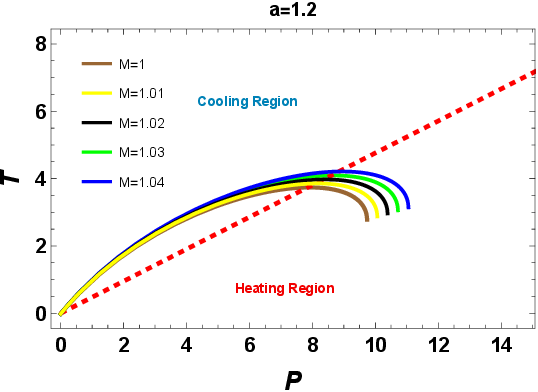}
\caption{\label{fa1}Isenthalpic and inversion plots in $T-P$ plane for
distinct choices of $a$ with $Q=0.65,\alpha =0.2$. }
\end{figure}

\begin{figure}[H]\centering
\includegraphics[width=.46\linewidth,height=2.6in]{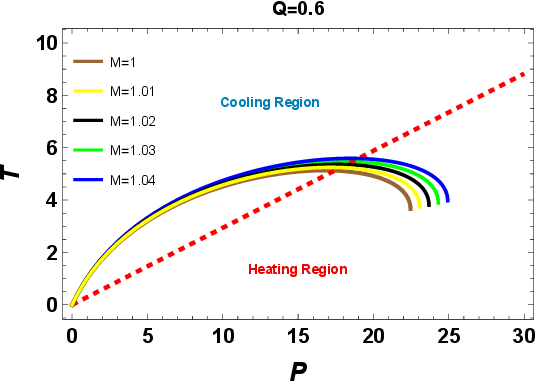}
\includegraphics[width=.46\linewidth,height=2.6in]{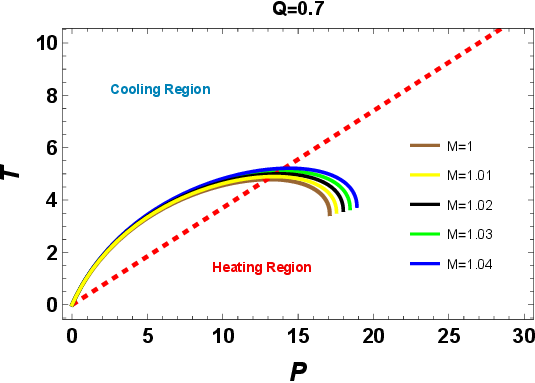}
\caption{\label{fa2}Isenthalpic and inversion plots in $T-P$ plane for
distinct choices of $Q$ with $a=0.6,\alpha =0.3$. }
\end{figure}

\begin{figure}[H]\centering
\includegraphics[width=.46\linewidth,height=2.6in]{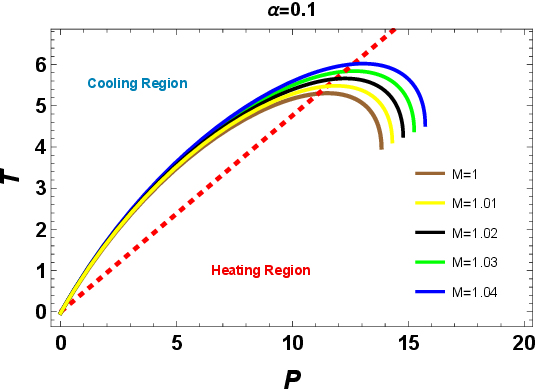}
\includegraphics[width=.46\linewidth,height=2.6in]{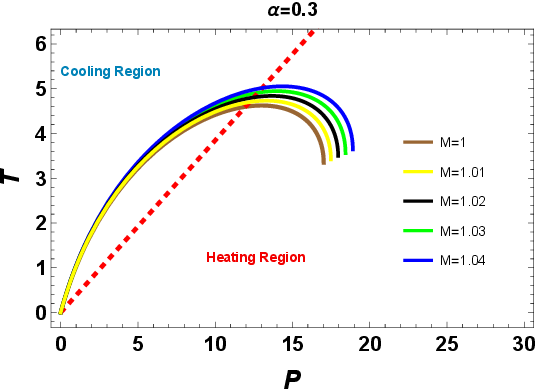}
\includegraphics[width=.46\linewidth,height=2.6in]{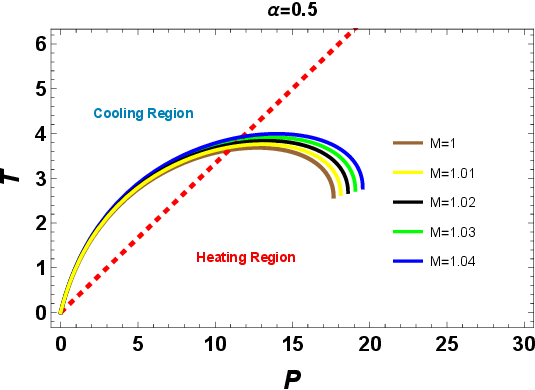}
\includegraphics[width=.46\linewidth,height=2.6in]{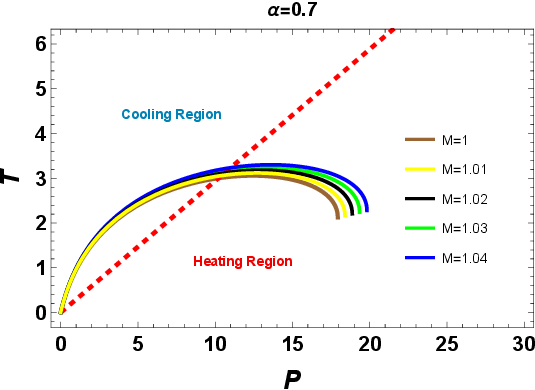}
\caption{\label{fa3}Isenthalpic and inversion plots in $T-P$ plane for
distinct choices of $\alpha$ with $Q=0.65,a =0.7$. }
\end{figure}

Starting with the condition $\mu_{J}=0$ in Eq.(\ref{3-2-3s}), one arrives at the following relation:
\begin{equation}\label{3-2-4}
4 \pi  r_{ci}^3 \left(a Q^4 \left(8 \alpha +5 \pi  r^2\right)-4 r_{ci}^4 \left(4 \alpha ^2+4 \alpha  Q^2+\pi  r_{ci}^2 \left(3 Q^2-2 r_{ci}^2\right)+(3 \pi -2) \alpha  r_{ci}^2\right)\right)=0.
\end{equation}
Under this framework, the parameter $P_i$ is interpreted as the inversion pressure. The valid solution of Eq. (\ref{3-2-4}), i.e., the positive real root, determines the associated horizon radius $r_{ci}$.

Furthermore, the explicit formula for the inversion temperature takes the form
\begin{eqnarray}\label{3-2-4ba}
T_i=-\frac{\left(a Q^4 \left(10 \alpha +7 \pi  r_{ci}^2\right)+4 r_{ci}^4 \left(\pi  r_{ci}^2 \left(-3 \alpha +48 \alpha  P_i r_{ci}^2-3 Q^2+r_{ci}^2\right)+8 \pi ^2 P_i r_{ci}^6-2 \alpha  \left(\alpha +Q^2+r_{ci}^2\right)\right)\right)}{48 r_{ci}^5 \left(2 \alpha +\pi  r_{ci}^2\right)^3\left(\pi  r_{ci}^2+4 \alpha  \log (r_{ci})\right)^{-1} }
\end{eqnarray}
Further, we can find 
\begin{eqnarray}\nonumber
T_i&=&-\frac{\left(\pi  r_{ci}^2+4 \alpha  \log (r_{ci})\right) \left(a Q^4+r_{ci}^4 \left(-2 \alpha -2 Q^2+r_{ci}^2\right)\right)}{4 r_{ci}^5 \left(6 \alpha ^2+\pi ^2 r_{ci}^4+3 \pi  \alpha  r_{ci}^2-2 \alpha  \left(6 \alpha +\pi  r_{ci}^2\right) \log (r)\right)},\\\nonumber P_i&=& \left(a Q^4 \left(\alpha +\pi  r_{ci}^2\right) \left(6 \alpha +5 \pi  r_{ci}^2\right)+2 \alpha  \log (r_{ci}) \left(a Q^4 \left(10 \alpha +7 \pi  r_{ci}^2\right)+4 \pi  r_{ci}^6 \left(-3 \alpha -3 Q^2+r_{ci}^2\right)\right.\right.\\\nonumber&-&\left.\left.8 \alpha  r_{ci}^4 \left(\alpha +Q^2+r_{ci}^2\right)\right)-4 r_{ci}^4 \left(6 \alpha ^2 \left(\alpha +Q^2-r_{ci}^2\right)+\pi  \alpha  r_{ci}^2 \left(7 \alpha +7 Q^2-5 r_{ci}^2\right)+\pi ^2 r_{ci}^4 \left(3 \alpha +3 Q^2-2 r_{ci}^2\right)\right)\right)\\\nonumber&\times&\left(32 \pi  r_{ci}^8 \left(6 \alpha ^2+\pi ^2 r_{ci}^4+3 \pi  \alpha  r_{ci}^2-2 \alpha  \left(6 \alpha +\pi  r_{ci}^2\right) \log (r_{ci})\right)\right)^{-1}
\end{eqnarray}
If one considers the special case $P_i=0$ in Eq. (\ref{3-2-4ba}), this reduces to the lowest attainable inversion temperature, given by
\begin{eqnarray}\nonumber
{T_{i}}^{min}&=&-\frac{\left(\pi  r_{ci}^2+4 \alpha  \log (r_{ci})\right) \left(a Q^4 \left(10 \alpha +7 \pi  r_{ci}^2\right)+4 \pi  r_{ci}^6 \left(-3 \alpha -3 Q^2+r_{ci}^2\right)-8 \alpha  r_{ci}^4 \left(\alpha +Q^2+r_{ci}^2\right)\right)}{48 r_{ci}^5 \left(2 \alpha +\pi  r_{ci}^2\right)^3}.
\end{eqnarray}
In Figures~(\ref{fa1})-(\ref{fa3}), we see the isenthalpic curves and inversion trajectories for the Joule–Thomson expansion of the charged Euler–Heisenberg–AdS BH in 4D EGB gravity in the \(T\)–\(P\) Plane. In all the figures, the inversion curve separates the thermodynamic phase space into the cooling (\(\mu_J>0\)) and heating (\(\mu_J<0\)) regions. This reinforces the idea of the Joule–Thompson expansion being nontrivial and similar to real fluids. The isenthalpic curves for different masses of the BH do intersect the inversion curve at different points. Beyond these points the nature of the thermodynamic response shifts. In Figure~(\ref{fa1}), we see how the parameter of the Euler–Heisenberg \(a\) affects Joule–Thomson expansion. When \(a\) increases, the Joule–Thomson inversion curve shifts to lower pressures and lower temperatures. This means strong nonlinear electromagnetic corrections limit the cooling region. Additionally, for larger \(a\), the inversion curve becomes steeper. This reflects that the sensitivity of BH temperature to changes in pressure increases. This shows how important NLED is in changing the thermodynamic phase structure of BHs. Figure~(\ref{fa2}) analyze the effects that the electric charge $Q$ has while the GB coupling and the nonlinear parameter are held constant. An increase in $Q$ raises both the inversion temperature and inversion pressure. This also increases the size of the cooling region in the $T - P$  plane. This is due to the increase in electromagnetic repulsion, which increases the thermal response of the BH when the expansion is isenthalpic. Therefore, the charged BHs will display more significant JT cooling effects than the weakly charged configurations. Figure (\ref{fa3}) shows how the JT expansion changes with the Gauss-Bonnet coupling \(\alpha\) . It appears that with the increase of \(\alpha\) the inversion curve moves to higher pressures and the inversion temperature slightly decreases, which shows that the corrections of high curvature are important to the expansion dynamics. The adjustments of \(\alpha\) comments on the effective equation of state which in turn changes the interplay of gravitational pull and thermodynamic pressure. It shows that the GB corrections are vital in the modifications of the heating and cooling transition of the BH system.

\section{Geodesic Equations}
\label{sec:5}
In gravitational theories, geodesic motion provides a direct link between
the geometry of spacetime and the dynamics of free test particles. The
trajectories of these particles encode information about the curvature and
symmetries of the underlying metric. In the framework of EGB gravity coupled
to EH NLED, this analysis becomes particularly relevant, since
the presence of additional fields modifies the spacetime geometry.
Consequently, the study of test particle motion offers an effective approach
to examine deviations from the predictions of general relativity induced by
EGB gravity and NLED. The investigation of geodesic motion is not limited to describing particle trajectories; it also provides insight into observable effects such as light
deflection, orbital dynamics, and photon propagation. Examining these trajectories enables one to understand the BH structure and the effects of a vector field on gravity. Now, we consider the Lagrangian to analyze the geodesic of considered BH given in Eq. (\ref{fun}) as
\begin{equation}
L=\frac{1}{2}g_{\mu \nu }\dot{x}^{\mu }\dot{x}^{\nu },
\end{equation}%
where $\dot{x}=dx/d\tau$. The metric from Eq. (\ref{fun}) allows the Lagrangian to be expressed as
\begin{equation}
\mathcal{L}=\frac{1}{2}\left[ r^{2}\left( \frac{d\theta }{d\tau }\right)
^{2}-\mathcal{F}\left( r\right) \left( \frac{dt}{%
d\tau }\right) ^{2}+\frac{1}{\mathcal{F}\left( r\right) }\left( \frac{dr}{%
d\tau }\right) ^{2}+r^{2}\sin ^{2}\theta \left( \frac{d\varphi }{d\tau }\right) ^{2}\right].
\end{equation}%
The Euler-Lagrange equations provide the principles of motion
\begin{equation}
\frac{d}{d\tau }\left( \frac{\partial }{\partial \dot{x}^{\sigma }}\mathcal{L%
}\right) -\left( \frac{\partial }{\partial x^{\sigma }}\mathcal{L}\right) =0.
\end{equation}%
Conservation of energy $E$ and angular momentum $L$ arise as conserved quantities because the metric does not depend on the coordinates $t$ and $\varphi$ respectively.
Mathematically, we can define
\begin{equation}
E=\mathcal{F}\left( r\right) \left( \frac{dt}{d\tau }\right) , 
\label{energy}
\end{equation}%
\begin{equation}
L=r^{2}\sin ^{2}\theta \left( \frac{d\varphi }{d\tau }\right) .\label{19}
\end{equation}%

These quantities correspond to conserved parameters associated with timelike
and axial Killing vectors and are measured by static observers at spatial
infinity. The equation governing the polar motion is obtained as
\begin{equation}
r^{2}\left( \frac{d^{2}\theta }{d\tau ^{2}}\right) +2r\left( \frac{dr}{d\tau 
}\right) \left( \frac{d\theta }{d\tau }\right) -r^{2}\sin \theta \cos \theta
\left( \frac{d\varphi }{d\tau }\right) ^{2}=0.
\end{equation}%
Fixing the Lagrangian as $\mathcal{L}=-\frac{\epsilon }{2}$, where  for
massive particles ($\epsilon =1$) and null geodesics ($\epsilon =0$), we obtain
\begin{equation}
\mathcal{F}\left( r\right) \left( \frac{dt}{d\tau }\right) ^{2}-\frac{1}{%
\mathcal{F}\left( r\right) }\left( \frac{dr}{d\tau }\right) ^{2}-r^{2}\left( 
\frac{d\theta }{d\tau }\right) ^{2}-r^{2}\sin ^{2}\theta \left( \frac{%
d\varphi }{d\tau }\right) ^{2}=\epsilon .
\end{equation}%
Using the conserved quantities, we get
\begin{equation}
\left( \frac{dr}{d\tau }\right) ^{2}=E^{2}-\mathcal{F}\left( r\right) \left(\epsilon +\frac{L^{2}}{r^{2} \sin ^{2}\theta}\right) .\label{54}
\end{equation}%
Alternatively, this equation takes the form%
\begin{equation}
\dot{r}^{2}=E^{2}-V_{\mathrm{eff}}\left( r\right) ,  \label{radial}
\end{equation}%
we obtain
\begin{equation}
V_{\mathrm{eff}}\left( r\right) =\mathcal{F}\left( r\right) \left( \epsilon +%
\frac{L^{2}}{r^{2} \sin ^{2}\theta}\right) .  \label{veff}
\end{equation}%
Eq. (\ref{energy}), Eq. (\ref{19}), Eq. (\ref{radial}) and Eq. (\ref%
{veff}) provide the basis for the analysis of geodesic motion in the
considered spacetime.

\subsection{Orbital geodesics of massive particles}

\begin{figure}[H]
\begin{minipage}[t]{0.5\textwidth}
        \centering
        \includegraphics[width=\textwidth]{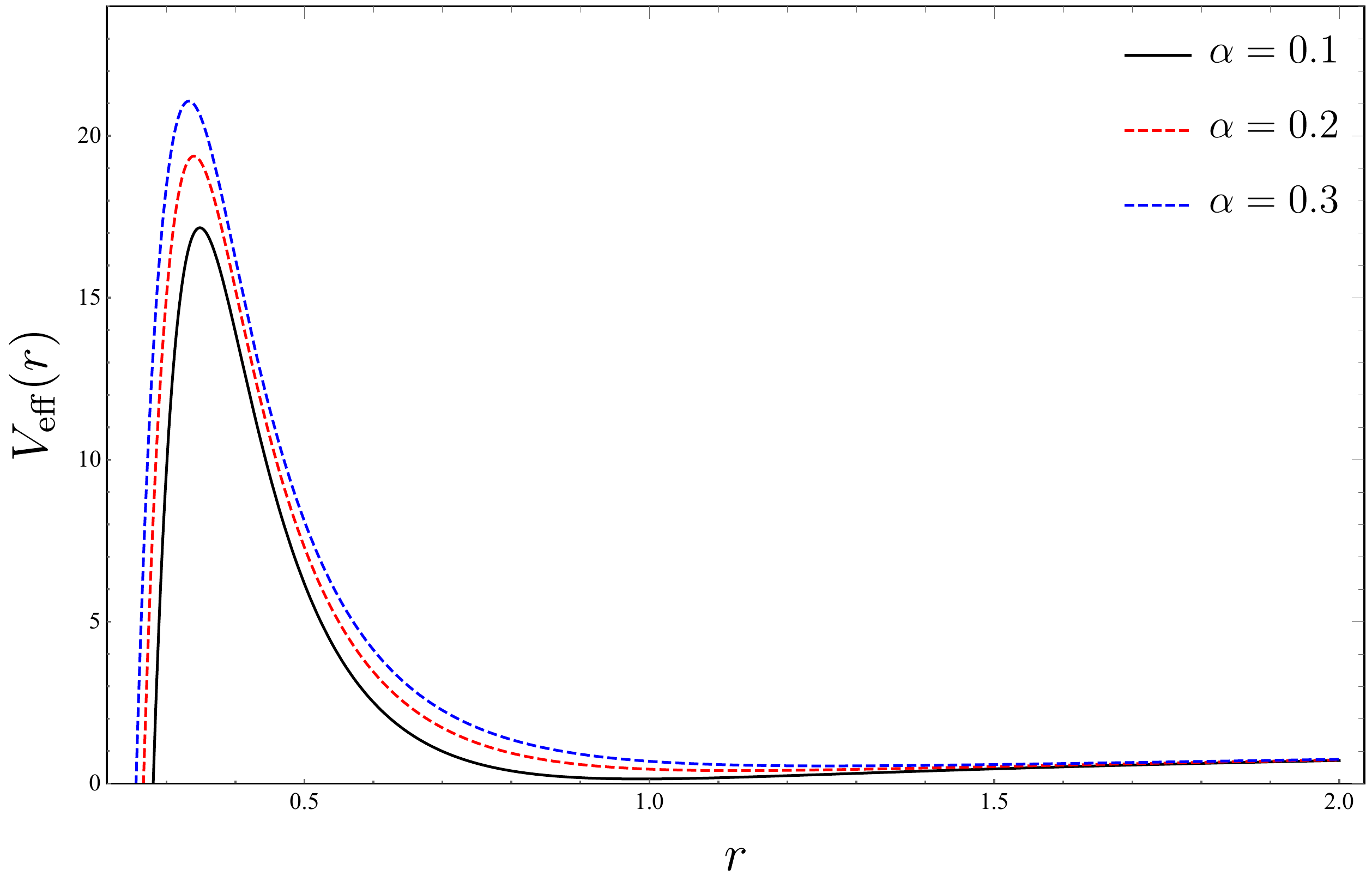}
                \subcaption{$a=0.1$ and $Q=0.9$}
        \label{fig:v1}
\end{minipage}
\begin{minipage}[t]{0.5\textwidth}
        \centering
        \includegraphics[width=\textwidth]{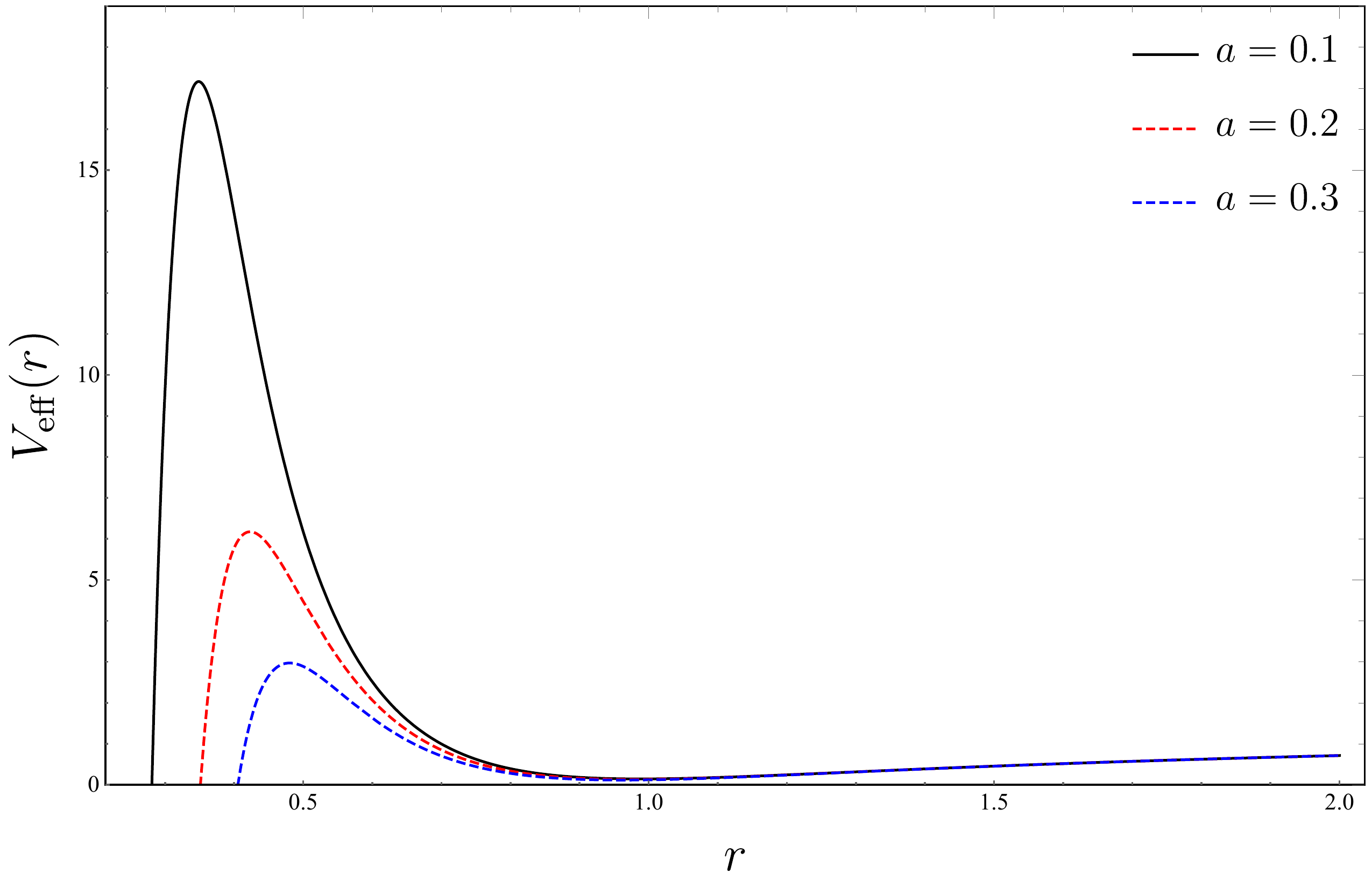}
                \subcaption{$\alpha=0.2$ and $Q=0.9$}
        \label{fig:v2}
\end{minipage}
\caption{Effective potential $V_{\mathrm{eff}}$ for massive test particles versus $r$ for $M=1$ and $\Lambda=-0.1$. }
\label{fig:Veff}
\end{figure}

Here, we investigate the geodesic motion of massive particles
with nonvanishing angular momentum $L\neq 0$ in the equatorial plane ($\theta =\frac{\pi }{2}$). For timelike geodesics ($%
\epsilon =1$), the effective potential takes the form%
\begin{equation}
V_{\mathrm{eff}}\left( r\right) =\left( 1+\frac{r^{2}}{2\alpha }\left( 1-%
\sqrt{1+\frac{4\alpha }{r^{2}}\left( \frac{2M}{r}-\frac{Q^{2}}{r^{2}}+\frac{%
\Lambda }{3}r^{2}+\frac{aQ^{4}}{20r^{6}}\right) }\right) \right) \left( 1+%
\frac{L^{2}}{r^{2}}\right) .
\end{equation}%
By varying $\alpha$ and $a$, we observe the behavior of $V_{\mathrm{eff}}\left( r\right) $ shown in Figure \ref{fig:Veff}. The detailed explanation is presented as:
\begin{itemize}
\item Panel (\ref{fig:v1}) of Fig.\ref{fig:Veff}  shows the effective potential for various values of
the Gauss--Bonnet coupling parameter. As $\alpha $ increases, the height of
the potential barrier increases and its maximum shifts slightly away from
the event horizon. This effect becomes less pronounced at larger radial
distances.

\item Panel (\ref{fig:v2}) displays the influence of the EH NLED
parameter. Increasing $a$ leads to a decrease in the maximum of the
effective potential, while the location of the peak moves outward,
indicating a modification of the orbital structure induced by NLED effects.

\end{itemize}

The effective potential determines the orbital properties of massive
particles in the considered spacetime. In general, the existence of local
minima in $V_{\mathrm{eff}}\left( r\right) $ corresponds to stable circular
orbits, while local maxima are associated with unstable circular
trajectories. The height and location of these extrema therefore play a
crucial role in determining the stability and admissible regions of particle
motion. The dependence of the effective potential on the GB coupling
parameter $\alpha $, indicates that higher-curvature corrections
significantly influence the orbital dynamics near the BH. As $\alpha 
$ increases, the enhancement of the potential barrier suggests that massive
particles require higher energy to access regions close to the event
horizon. However, the decreasing influence of $\alpha $ at larger radial
distances implies that GB effects are primarily relevant in the strong-field
regime, while the spacetime gradually approaches the general relativistic
behavior far from the BH. The contribution of the EH
NLED parameter $a$ leads to a noticeable modification of the effective
potential. The reduction in the height of the potential maximum with
increasing $a$ indicates a weakening of the repulsive contribution
associated with nonlinear electromagnetic effects. This behavior suggests
that NLED may facilitate the inward motion of massive
particles and alter the conditions for stable circular orbits when compared
to the linear Maxwell case. Overall, these results demonstrate that both
higher-curvature corrections and nonlinear electrodynamic effects introduce
nontrivial modifications to the geodesic structure of the spacetime. The
combined influence of these parameters leads to significant deviations from
the standard scenarios, particularly in the vicinity of the BH,
thereby affecting the existence and stability of massive particle orbits.
We now examine the motion of test particles along nonradial timelike
geodesics by focusing on their orbital dynamics. The nature of the resulting trajectories is determined by the value of the conserved energy $E$, allowing the motion to be classified into several distinct regimes:
\begin{description}
\item[$E>E_{\mathrm{uns}}$:] The absence of radial turning points implies that particles are inevitably captured by the BH. In this regime, the gravitational attraction dominates over the centrifugal barrier, resulting in plunge-type trajectories. Such orbits are characteristic of the strong-field region and are particularly sensitive to higher-curvature and nonlinear electrodynamic corrections.

\item[$E=E_{\mathrm{uns}}$:] The particle occupies an unstable circular orbit
located at the maximum of the effective potential. This configuration
represents the threshold between plunging and escaping trajectories. Any
small perturbation causes the particle either to fall into the BH or
to move away toward large radial distances, reflecting the marginal stability
of this orbit.

\item[$E=E_{3}$:]  Particles approach the BH from large distances,
reach a minimum radial position, and are subsequently deflected back. This
regime corresponds to scattering-type trajectories and is directly related
to gravitational deflection phenomena in the considered spacetime.

\item[$E=E_{2}$ :] The effective potential exhibits two turning points,
allowing for bound motion. In this case, particles oscillate between
perihelion and aphelion distances, forming stable bound orbits governed by a
balance between gravitational attraction and centrifugal effects.

\item[$E=E_{\mathrm{sta}}$:-] The particle follows a stable circular orbit
located at the minimum of the effective potential. Small radial perturbations
about this radius lead to bounded oscillations, indicating local stability.
\end{description}

These results show that the energy dependent structure of timelike geodesics
is significantly modified by GB and EH corrections
in the strong-field regime, while the standard behavior
is recovered at larger radial distances.

\begin{figure}[H]
    \centering
    \begin{subfigure}{.65\textwidth}
        \centering
        \includegraphics[width=1.0\linewidth]{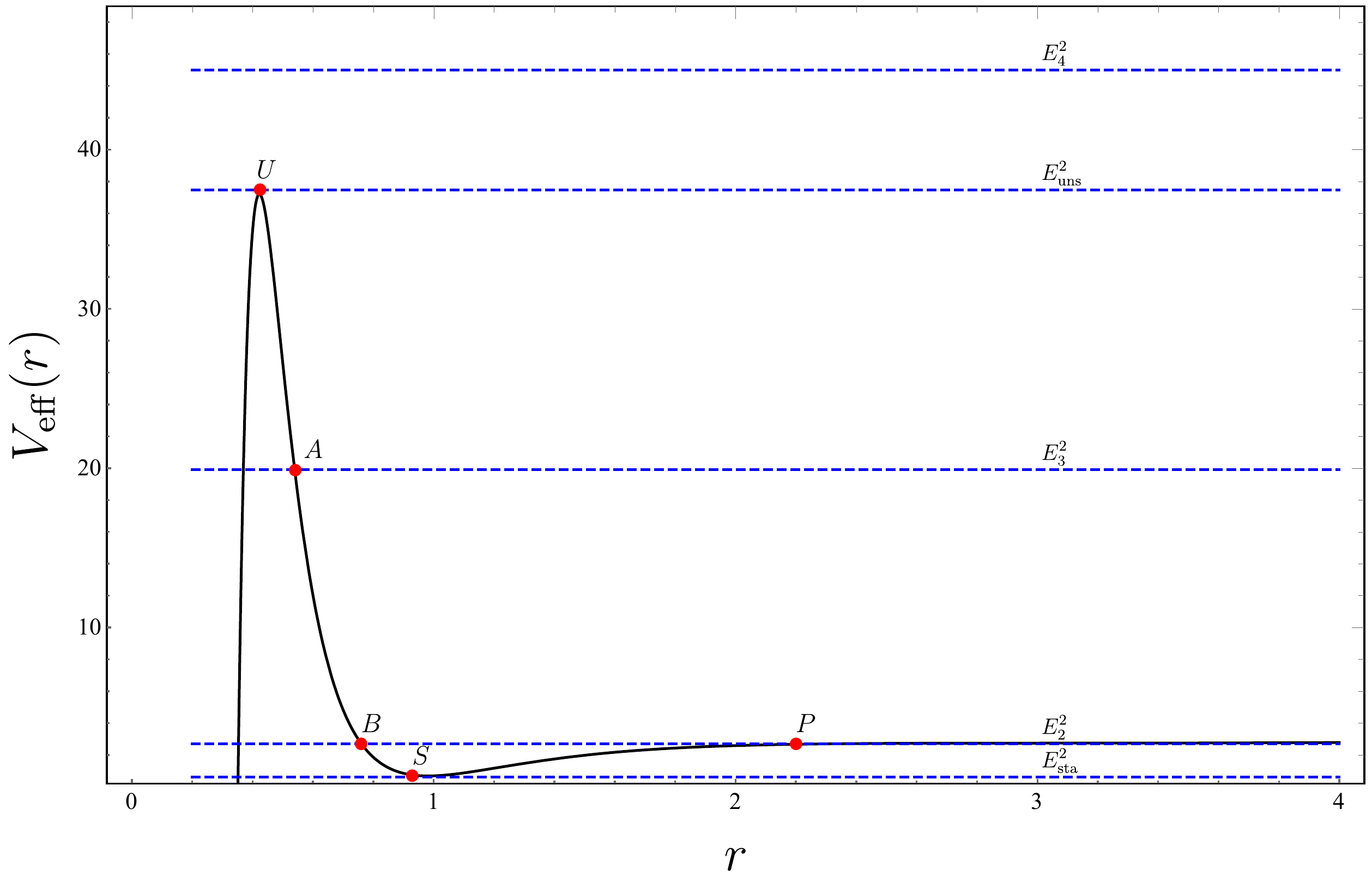}
        \caption{Effective potential}
        \label{sub1}
    \end{subfigure}
    \vspace{-2mm}
    \begin{subfigure}{.37\textwidth}
        \centering
        \includegraphics[width=1.0\linewidth]{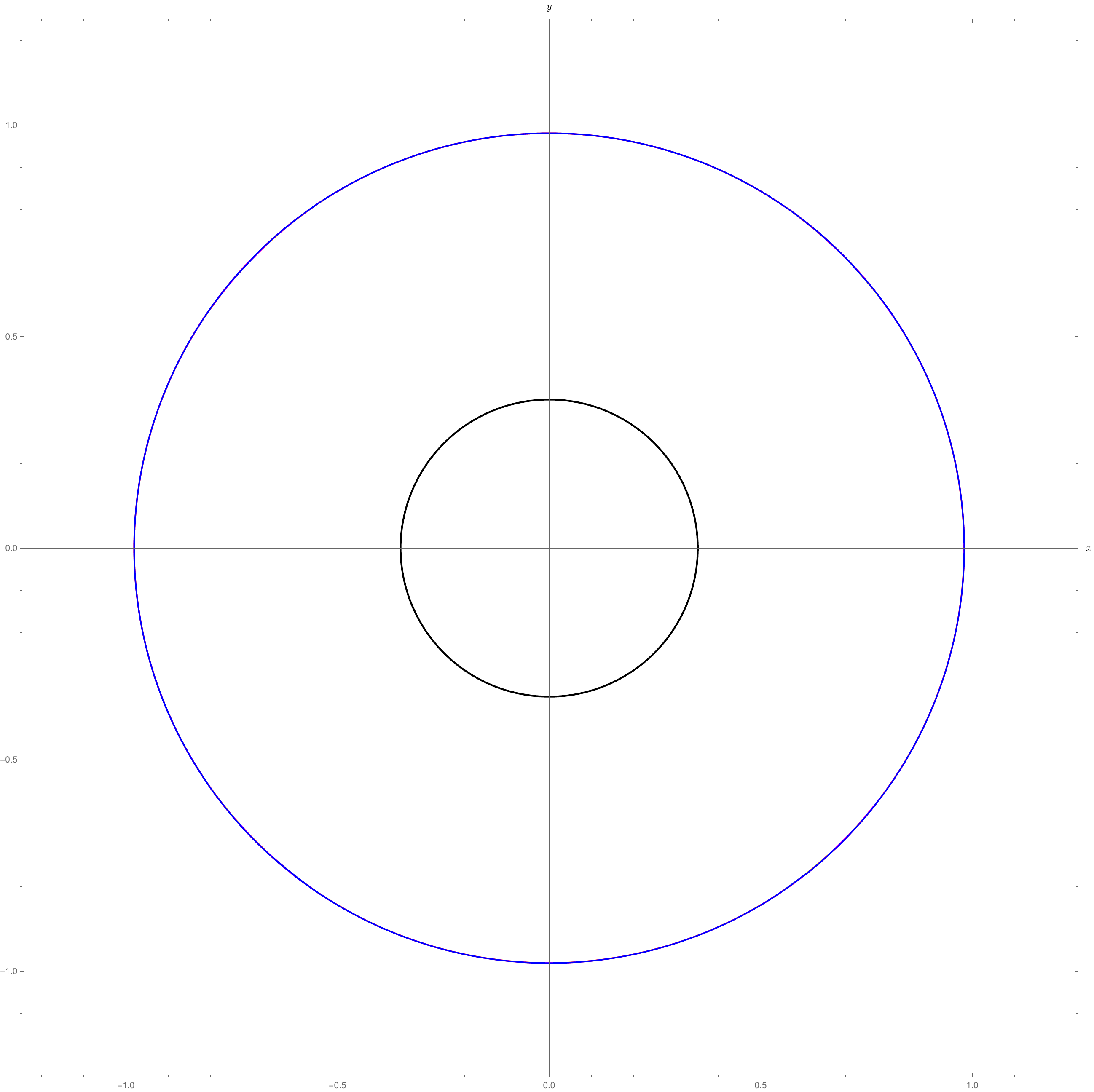}
        \caption{stable circular orbit: $E=E_{\mathrm{sta}}$}
        \label{sub2}
    \end{subfigure}
    \hfill
    \begin{subfigure}{.37\textwidth}
        \centering
        \includegraphics[width=1.0\linewidth]{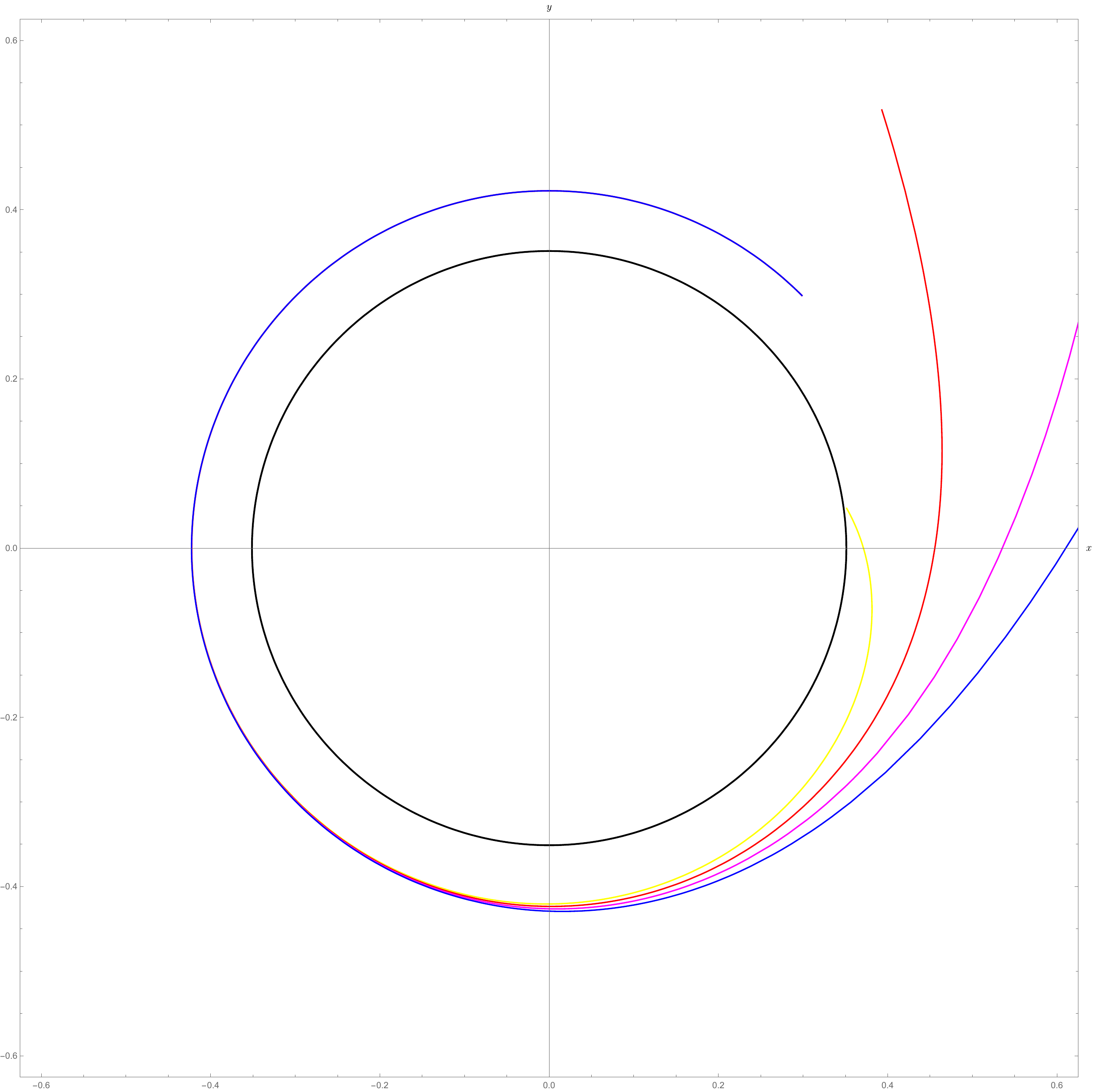}
        \caption{unstable circular orbit: $E=E_{\mathrm{uns}}$}
        \label{sub3}
    \end{subfigure}
     \vspace{-2mm}
      \begin{subfigure}{.37\textwidth}
        \centering
        \includegraphics[width=1.0\linewidth]{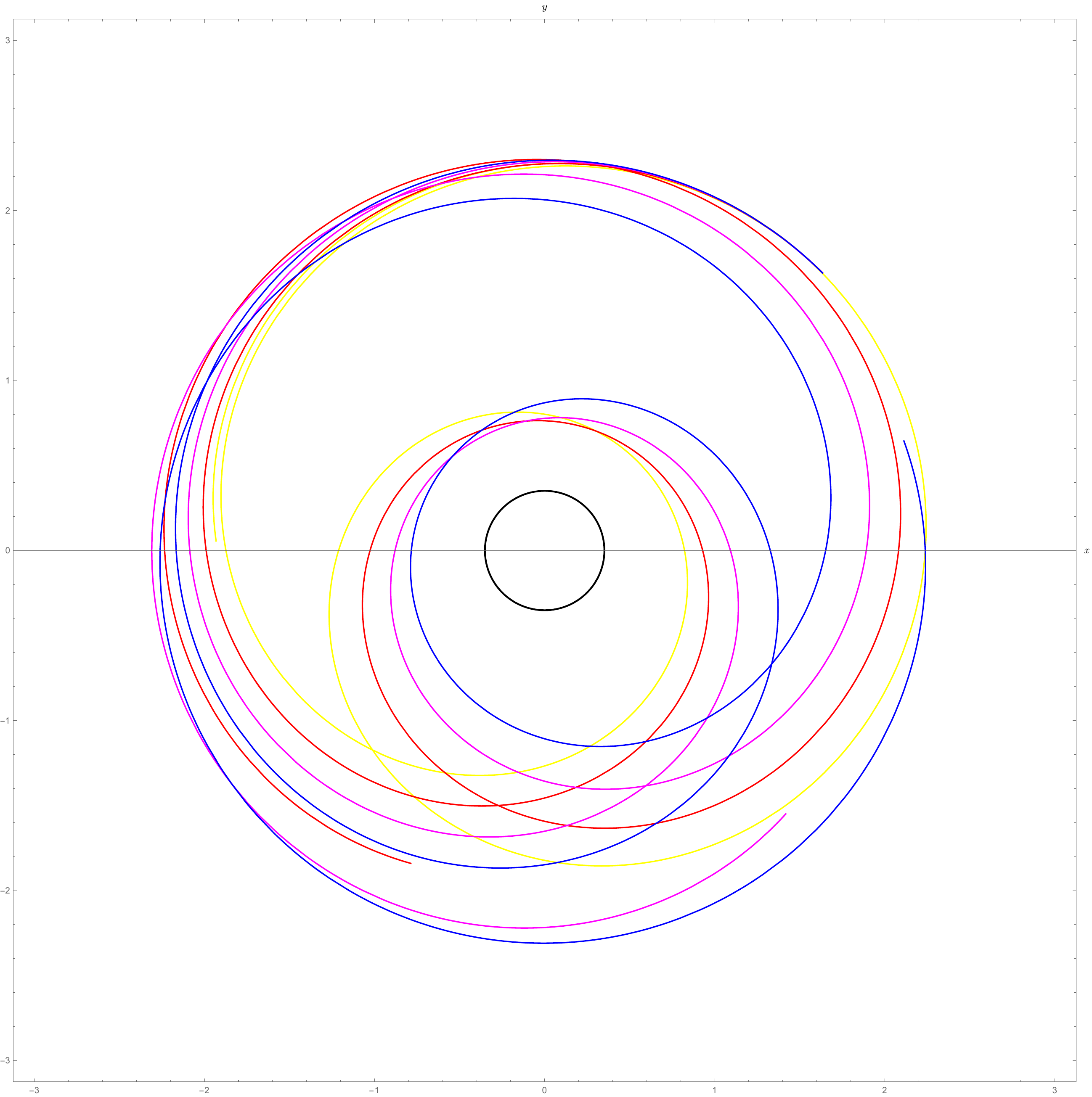}
        \caption{plunging trajectory: $E=E_{2}$}
        \label{sub4}
    \end{subfigure}
    \hfill
    \begin{subfigure}{.37\textwidth}
        \centering
        \includegraphics[width=1.0\linewidth]{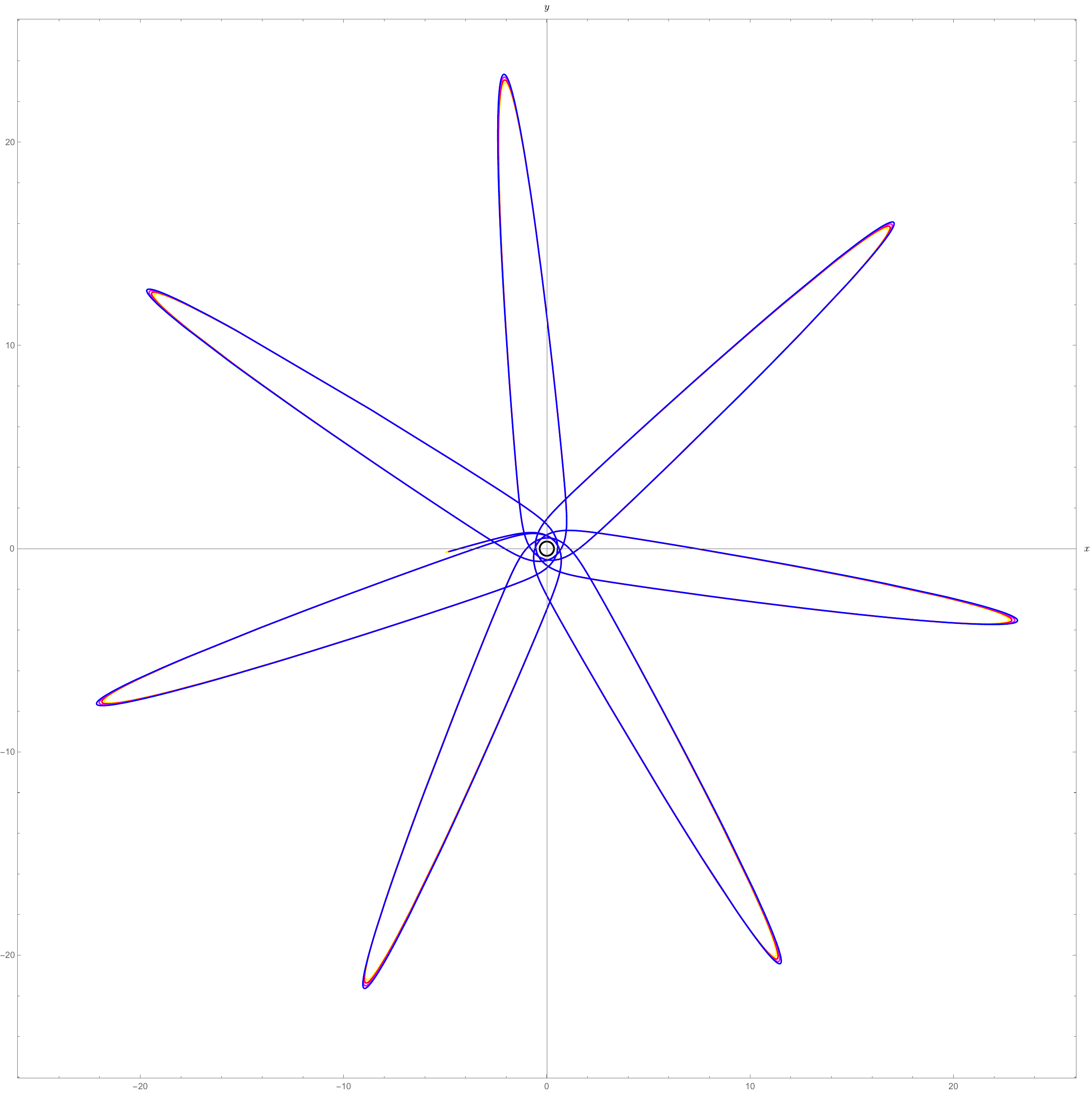}
        \caption{bound orbit with two turning points:$E=E_{3}$}
        \label{sub5}
    \end{subfigure}
    \caption{Effective potential and possible timelike particle motions in the 4D EGB–Euler–Heisenberg BH spacetime. Parameters: $M=1$, $Q=0.9$, $a=0.2$, $\alpha=0.2$, $\Lambda=-0.1$ and $L=4$.}
    \label{fig:ueff}
\end{figure}
In the following, we investigate circular trajectories and their stability
in order to assess how the deformed spacetime geometry affects this class of
motion. Circular orbits are characterized by the simultaneous conditions
that the radial velocity and radial acceleration vanish, namely 
\begin{equation}
\dot{r}=\ddot{r}=0,
\end{equation}%
and the corresponding effective potential must satisfy%
\begin{equation}
V_{\mathrm{eff}}\left( r\right) =E^{2},\frac{d}{dr}V_{\mathrm{eff}}\left(
r\right) =0.  \label{18}
\end{equation}%
The stability of circular geodesics is determined by the effective potential behavior in the vicinity of the circular orbit radius. A circular
orbit is stable if the effective potential attains a local minimum, which
requires%
\begin{equation}
\left. \frac{d^{2}}{dr^{2}}V_{\mathrm{eff}}\left( r\right) \right\vert
_{r=r_{\mathrm{sta}}}>0,
\end{equation}%
whereas a local maximum,%
\begin{equation}
\left. \frac{d^{2}}{dr^{2}}V_{\mathrm{eff}}\left( r\right) \right\vert
_{r=r_{\mathrm{uns}}}<0.
\end{equation}%
Numerical solutions of Eq. (\ref{18}) are presented in Table \ref{tab:stab}, where the
radii of stable and unstable circular orbits are listed for various values
of the GB coupling $\alpha $ and the EH parameter $a$. The
table illustrates how the locations of both stable and unstable circular
orbits depend on the $\alpha $ and $a$. The numerical results indicate that
the GB coupling has a significant impact on orbital stability. As 
$\alpha $ increases, the radius of the stable circular orbit shifts toward
larger values, while the corresponding unstable orbit moves inward. This
trend suggests that higher-curvature corrections tend to enhance the range
of stable circular motion for massive particles. By contrast, variations in
the EH  parameter produce an
opposite effect. Increasing $a$ results in a reduction of the radius of
stable circular orbits, accompanied by an outward displacement of unstable
circular orbits. This behavior reflects the role of NLED effects in altering
the interplay between gravitational attraction and centrifugal forces,
thereby narrowing the region of stable circular motion.

\begin{table}[H]
\centering
\begin{tabular}{|c|c|c||c|c|c|}
\hline
\multicolumn{3}{|c||}{$a=0.1$} & \multicolumn{3}{c|}{$\alpha =0.2$} \\ \hline
$\alpha $ & $r_{\mathrm{sta}}$ & $r_{\mathrm{uns}}$ & $a$ & $r_{\mathrm{sta}%
} $ & $r_{\mathrm{uns}}$ \\ \hline
0.1 & 0.81657 & 0.360482 & 0.1 & 0.939702 & 0.348674 \\ 
0.2 & 0.939702 & 0.348674 & 0.2 & 0.926724 & 0.423514 \\ 
0.3 & 1.04635 & 0.339858 & 0.3 & 0.912094 & 0.481236 \\ 
0.4 & 1.13755 & 0.332581 & 0.4 & 0.895157 & 0.53199 \\ 
0.5 & 1.21435 & 0.326257 & 0.5 & 0.874697 & 0.580648 \\ \hline
\end{tabular}%
\caption{Radii of stable and unstable circular orbits. Parameters $M=1$, $%
Q=0.9$, $\Lambda =-0.6$ and $L=3.$.}
\label{tab:stab}
\end{table}
The angular velocity of test particles provides a direct probe of the
spacetime geometry and allows us to quantify the influence of the GB and
NLED parameters on orbital motion. The angular velocity of a particle
orbiting a BH, as measured by an observer at infinity and commonly
referred to as the Keplerian frequency, is defined as%
\begin{equation}
\Omega _{K}=\frac{d\varphi }{dt}=\frac{\dot{\varphi}}{\dot{t}}=\sqrt{\frac{%
\mathcal{F}\left( r\right) ^{\prime }}{2r}}.
\end{equation}

\begin{figure}[H]
\begin{minipage}[t]{0.5\textwidth}
        \centering
        \includegraphics[width=\textwidth]{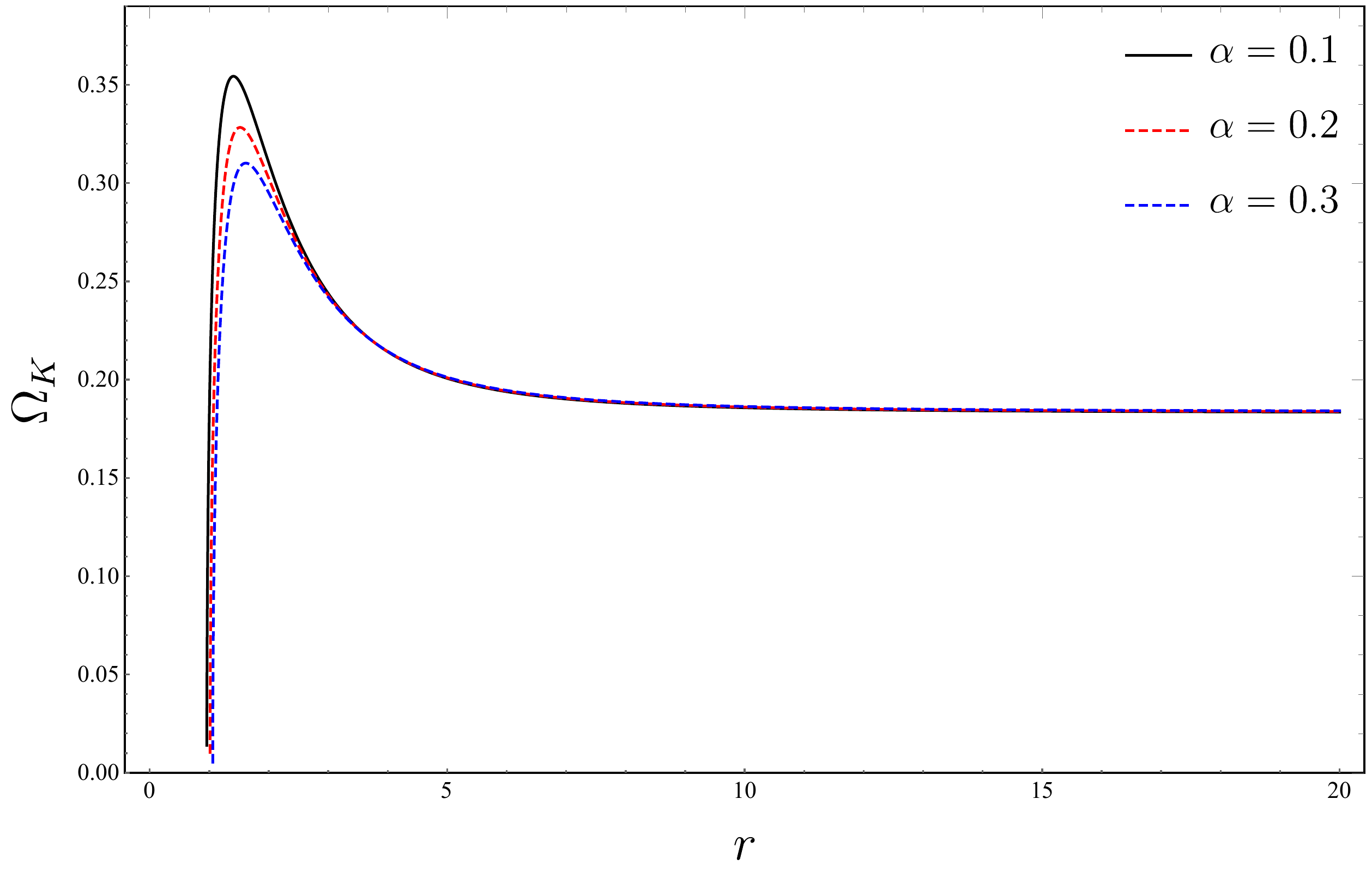}
                \subcaption{$a=0.1$ and $Q=0.9$}
        \label{fig:K1}
\end{minipage}
\begin{minipage}[t]{0.5\textwidth}
        \centering
        \includegraphics[width=\textwidth]{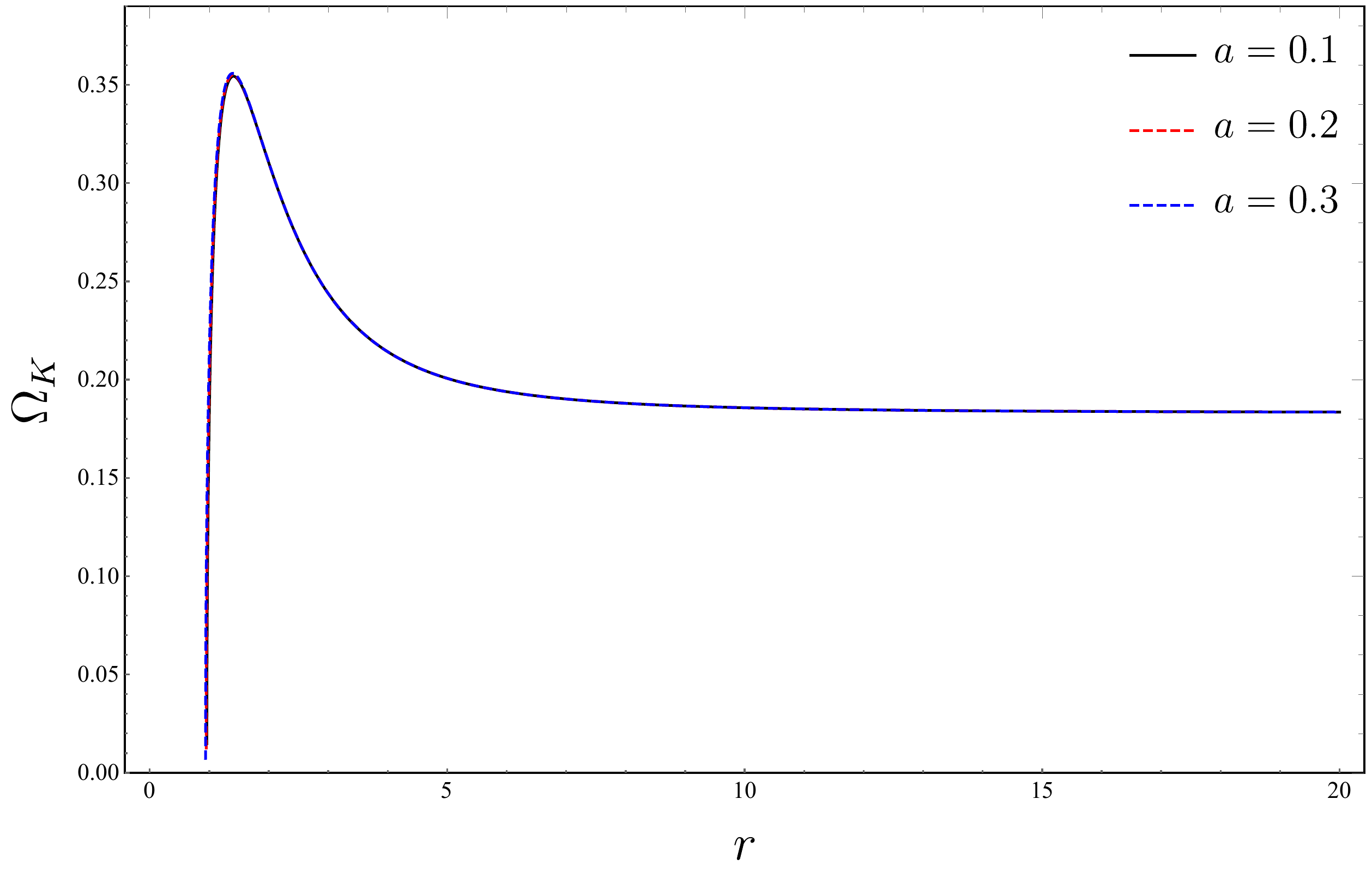}
                \subcaption{$\alpha=0.2$ and $Q=0.9$}
        \label{fig:K2}
\end{minipage}
\caption{Keplerian frequency $\Omega_{K}$ for massive test particles versus $r$ for $M=1$ and $\Lambda=-0.1$. }
\label{fig:Kepler}
\end{figure}

\begin{figure}[H]
\begin{minipage}[t]{0.5\textwidth}
        \centering
        \includegraphics[width=\textwidth]{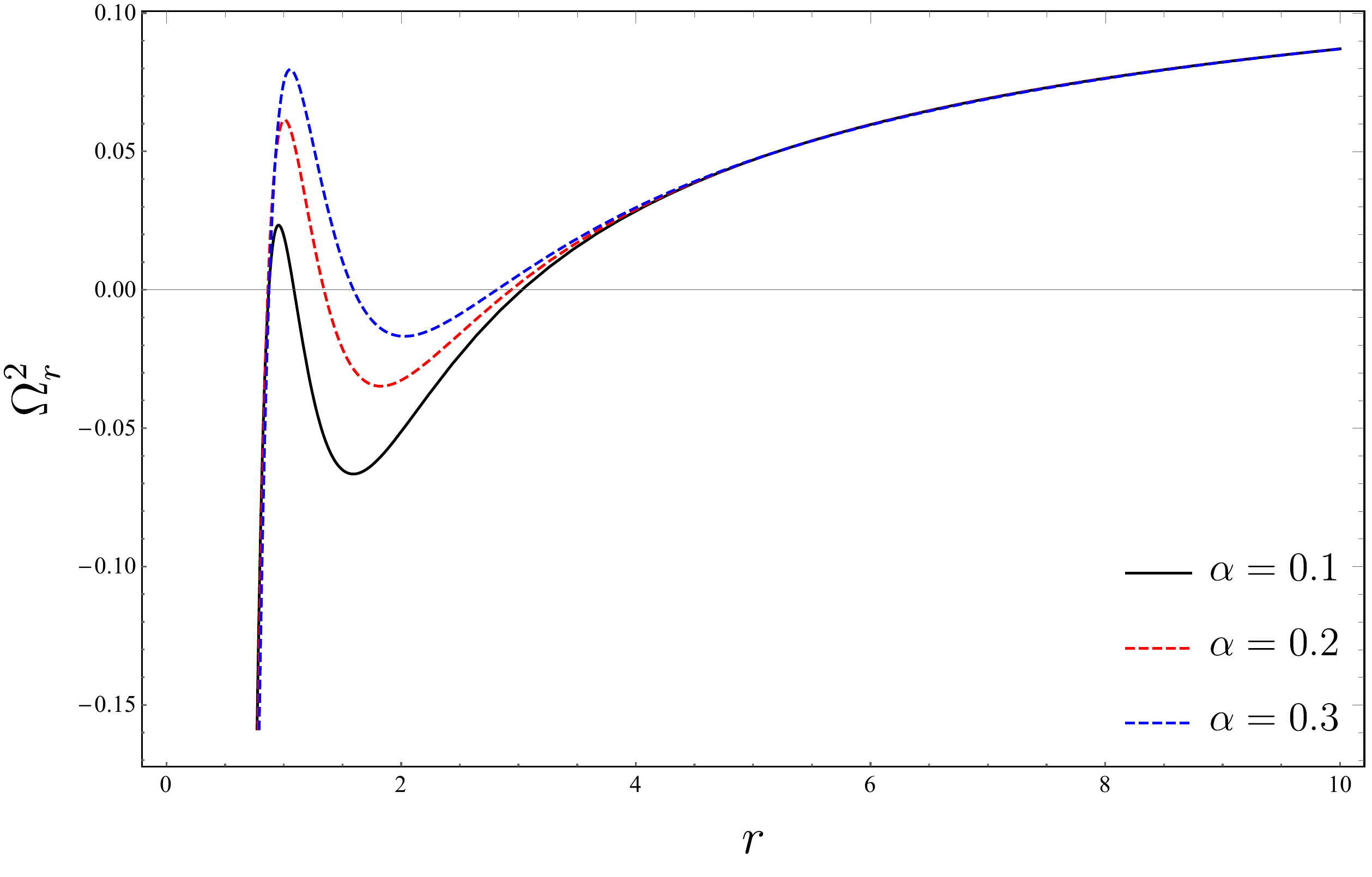}
                \subcaption{$a=0.1$ }
        \label{fig:o1}
\end{minipage}
\begin{minipage}[t]{0.5\textwidth}
        \centering
        \includegraphics[width=\textwidth]{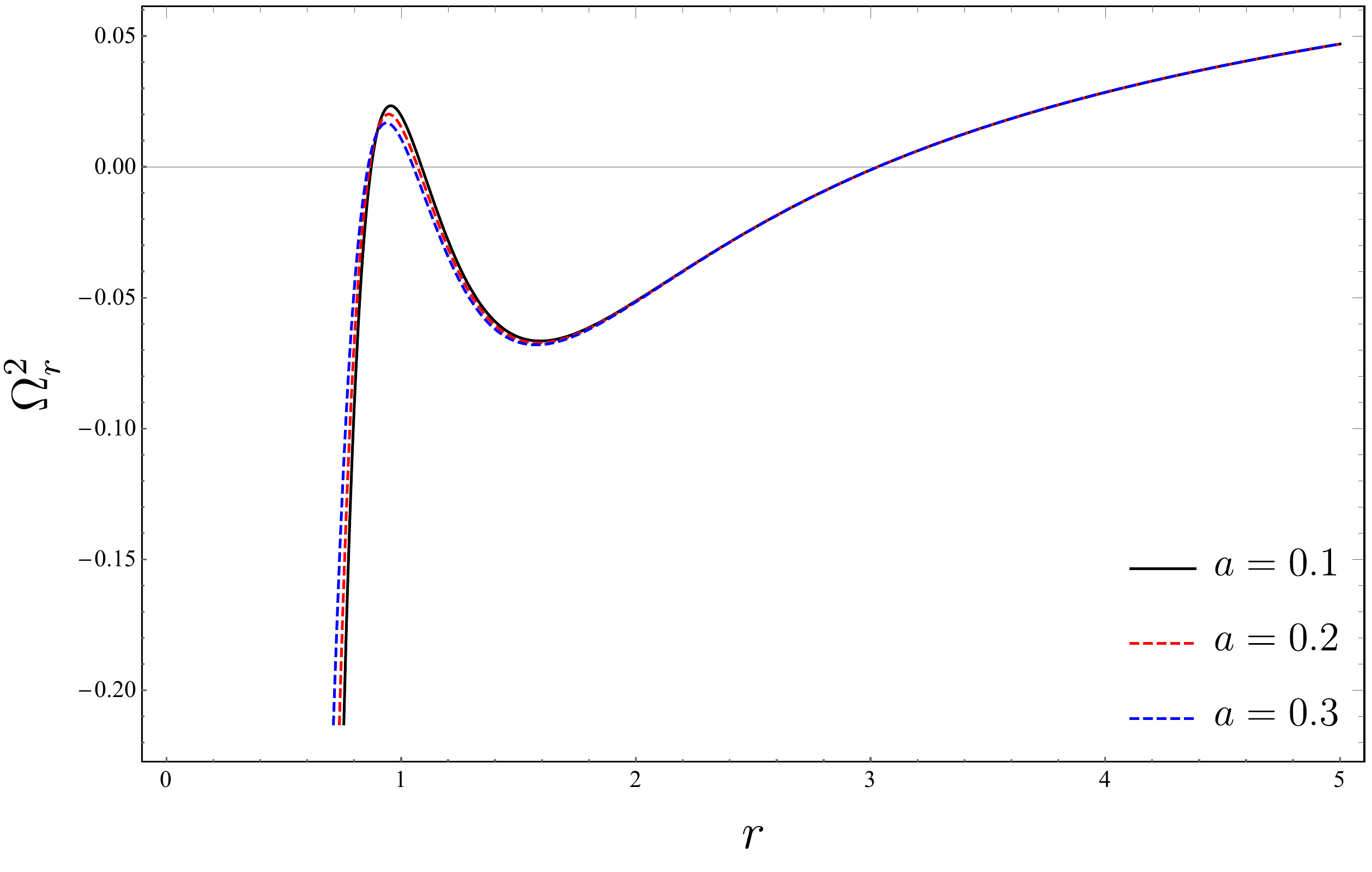}
                \subcaption{$\alpha=0.2$}
        \label{fig:o2}
\end{minipage}
\caption{Radial profiles of the epicyclic frequencies $\Omega^2_{r}$  versus $r$ for $M=1$ and $\Lambda=-0.1$. }
\label{fig:mega}
\end{figure}

The radial dependence of the Keplerian frequency for test particles orbiting
the EH BH in EGB gravity is shown in Fig.\ref{fig:Kepler}. One
observes that the Keplerian frequency initially increases with the radial
coordinate, reaches a maximum at an intermediate radius, and then decreases
before asymptotically approaching a constant value at large distances.
Furthermore, increasing the GB coupling parameter $\alpha $ leads to a
reduction in the maximum value of the Keplerian frequency, indicating that
higher-curvature corrections weaken the orbital angular velocity near the
BH.
To investigate the orbital dynamics of test particles, we consider small oscillations around stable circular equatorial orbits. These oscillatory motions are characterized by the radial and vertical epicyclic frequencies, which provide important information about orbital stability and the influence of spacetime curvature. The stability of circular motion in the radial and vertical directions can be analyzed through the corresponding epicyclic frequencies $\Omega _{r}$ and $\Omega _{\theta }$ . From Eqs. (\ref{energy}) and (\ref{54}), small radial and polar perturbations about a circular equatorial orbit satisfy

\begin{equation}
\frac{1}{2}\left( \frac{\dot{r}}{\dot{t}}\right) ^{2}=\frac{1}{2}\left( 
\frac{dr}{dt}\right) ^{2}=-\frac{1}{2}\frac{\mathcal{F}\left( r\right) ^{3}}{%
E^{2}}\left[ 1-\frac{E^{2}}{\mathcal{F}\left( r\right) }+\frac{L^{2}}{%
r^{2}\sin ^{2}\theta }\right] \equiv V_{\mathrm{eff}}^{\left( r\right) },
\label{315}
\end{equation}%
\begin{equation}
\frac{1}{2}\left( \frac{\dot{\theta}}{\dot{t}}\right) ^{2}=\frac{1}{2}\left( 
\frac{d\theta }{dt}\right) ^{2}=\frac{1}{2}\frac{\mathcal{F}\left( r\right)
^{2}}{r^{2}E^{2}}\left[ 1-\frac{E^{2}}{\mathcal{F}\left( r\right) }+\frac{%
L^{2}}{r^{2}\sin ^{2}\theta }\right] \equiv V_{\mathrm{eff}}^{\left( \theta
\right) },  \label{316}
\end{equation}%
where the $\sin \theta $ factor is retained to account for polar orbital perturbations. Introducing small deviations $\delta r$ and $\delta \theta $,
and differentiating Eqs. (\ref{315}) and (\ref{316}), with respect to the coordinate time $t$, we obtain

\begin{eqnarray}
\frac{d^{2}}{dt^{2}}\delta r &=&\frac{d^{2}V_{\mathrm{eff}}^{\left( r\right)
}}{dr^{2}}\delta r, \\
\frac{d^{2}}{dt^{2}}\delta \theta  &=&\frac{d^{2}V_{\mathrm{eff}}^{\left(
\theta \right) }}{d\theta ^{2}}\delta \theta .
\end{eqnarray}%
The radial and vertical epicyclic frequencies are therefore given by
\begin{equation}
\Omega _{r}^{2}=\frac{d^{2}V_{\mathrm{eff}}^{\left( r\right) }}{dr^{2}}\text{
\ \ and  }\Omega _{\theta }^{2}=\frac{d^{2}V_{\mathrm{eff}}^{\left( \theta
\right) }}{d\theta ^{2}}.
\end{equation}%
Combining Eqs. (\ref{18}), (\ref{315}) and (\ref{316}), the explicit expressions for the epicyclic frequencies read
\begin{equation}
\Omega^2 _{r}=-\mathcal{F}^{\prime }\left( r\right) ^{2}+\frac{\mathcal{F%
}\left( r\right) }{2r}\left[ 3\mathcal{F}^{\prime }\left( r\right) +r%
\mathcal{F}^{\prime \prime }\left( r\right) \right] ,
\end{equation}%
\begin{equation}
\Omega^2 _{\theta }=\frac{%
\mathcal{F}\left( r\right) ^{\prime }}{2r}.
\end{equation}
Figure \ref{fig:mega} shows the radial profiles of the epicyclic frequencies $\Omega^2_{r}$ as functions of the radial coordinate $r$. From Fig.\ref{fig:mega}, one observes that $\Omega^2_{r}$ becomes negative over a finite radial interval, indicating the presence of radially unstable circular orbits. The radial epicyclic frequency squared exhibits three zeros, separating regions of stable and unstable circular motion, with the outermost zero corresponding to the physically relevant ISCO. Furthermore, increasing GB coupling $\alpha$ shifts the zero of $\Omega^2_{r}$ toward larger values of $r$ implying that higher-curvature corrections enlarge the unstable region and move the ISCO outward. In contrast, increasing the EH parameter $a$ shifts the zero of the radial epicyclic frequency toward smaller values of the radial coordinate $r$, indicating that NLED effects reduce the radius of the ISCO.
The ISCO is of particular importance in astrophysics, as it defines the inner edge of accretion disks and sets a reference point in the inspiral phase of compact binaries, thereby influencing the emitted gravitational-wave signal and providing information about the strong-field region of the spacetime. This class of orbits is determined by the stability condition, which is given by
\begin{equation}
\Omega^2 _{r_{\mathrm{ISCI}}}= 0.   
\end{equation}
Due to the complexity of the metric function, the ISCO radius is evaluated numerically for different values of the model parameters, and the results are presented in Table \ref{tab:ISCO}. The table shows the dependence of the ISCO parameters ($r_{\mathrm{ISCO}}$, $L_{\mathrm{ISCO}}$, $E_{\mathrm{ISCO}}$) depend on $\alpha $ and $a$. The results indicate that the GB coupling strongly influences the ISCO, reducing the radius, angular momentum, and energy, and allowing stable orbits closer to the horizon. By contrast, variations in the EH parameter, produces only minimal changes in ISCO parameters, indicating that GB corrections dominate the orbital dynamics, while NLED effects are comparatively negligible.
\begin{table}[H]
\centering
\begin{tabular}{|c|c|c|c||c|c|c|c|}
\hline
\multicolumn{4}{|c||}{$a=0.1$} & \multicolumn{4}{c|}{$\alpha =0.2$} \\ \hline
$\alpha $ & $r_{\mathrm{ISCO}}$ & $L_{\mathrm{ISCO}}$ & $E_{\mathrm{ISCO}}$
& $a$ & $r_{\mathrm{ISCO}}$ & $L_{\mathrm{ISCO}}$ & $E_{\mathrm{ISCO}}$ \\ 
\hline
0.1 & 3.10561 & 5.10453 & 1.67953 & 0.1 & 3.02981 & 5.03499 & 1.66349 \\ 
0.2 & 3.02981 & 5.03499 & 1.66349 & 0.2 & 3.03009 & 5.03526 & 1.66356 \\ 
0.3 & 2.93708 & 4.95006 & 1.64376 & 0.3 & 3.03037 & 5.03553 & 1.66362 \\ 
0.4 & 2.81557 & 4.84202 & 1.61841 & 0.4 & 3.03064 & 5.03581 & 1.66368 \\ 
0.5 & 2.62659 & 4.69174 & 1.58259 & 0.5 & 3.03092 & 5.03608 & 1.66374 \\ 
\hline
\end{tabular}
\caption{ISCO parameters for test particles for various possible scenarios.}
\label{tab:ISCO}
\end{table}
\section{Conclusions}

\label{sec:6}
In this work, we have investigated charged AdS BH solutions in 4D EGB gravity coupled to EH NLED. By employing the regularized 4D EGB framework and the weak-field limit of the EH effective Lagrangian, we constructed exact BH solutions and explored the combined effects of higher-curvature corrections and nonlinear electromagnetic interactions on the geometric, thermodynamic, and dynamical properties of the spacetime.
We first analyzed the horizon structure of the BH and showed that the presence of the GB coupling and the EH parameter significantly modifies the number and location of horizons. The nonlinear electromagnetic corrections introduce higher-order charge contributions, while the GB term alters the short-distance behavior of the metric, leading to notable deviations from the standard Reissner-Nordström-AdS geometry.

Within the extended phase space formalism, where the cosmological constant is interpreted as thermodynamic pressure, we examined the thermodynamic behavior of the system in detail. We derived the mass, Hawking temperature, entropy, and heat capacity, highlighting the role of the GB coupling in generating logarithmic corrections to the entropy. The thermodynamic analysis revealed the existence of critical behavior analogous to the liquid--gas phase transition, with the GB term playing a dominant role in shifting the critical point, while the EH parameter produces comparatively milder corrections. The heat capacity analysis further confirmed the presence of second-order phase transitions separating thermodynamically stable and unstable BH branches.

The JT expansion was studied by interpreting the BH mass as enthalpy, which allows us to obtain Joule Thomson coefficient, inversion temperature, and inversion pressure of an isenthalpic BH expansion. The presence of a nontrivial inversion curve indicates that BH is undergoing both heating and cooling, similar to a real thermal system. NLED effects that are described by the EH parameter a, greatly reduce the cooling region of the BH which is characterized by a low inversion temperature and pressure. Hence, these results are an indication of the non-maxwellian behavior of electrodynamics. Charged $Q$ markedly influences the JT Effect by significantly increasing the inversion temperature and broadening the cooling region of the BH within the temperature $T$ and pressure $P$ coordinates. On the contrary, the GB coupling $\alpha$ which is associated with higher order curvature corrections, modifies the thermodynamic structure by changing the location of the inversion curve and the interplay of gravity and pressure.

Finally, we investigated the geodesic motion of test particles in the BH spacetime. The analysis of effective potentials, circular orbits, Keplerian frequencies, and orbital stability demonstrated that GB and EH corrections significantly affect particle dynamics in the strong-field regime. Higher-curvature effects tend to enlarge the region of stable circular orbits, while nonlinear electromagnetic corrections reduce it, leading to a nontrivial interplay between gravitational and electromagnetic contributions.

These results indicate that the combined effects of EGB gravity and EH NLED lead to rich and physically meaningful modifications of BH properties. This study provides further insight into the role of higher-curvature gravity and QED corrections in BH physics. Possible extensions of this work include the analysis of quasinormal modes, BH shadows, and rotating solutions within the same framework, which may offer additional observational signatures of these corrections.

\end{document}